\documentclass[%
 aip,
rsi,%
 amsmath,amssymb,
 reprint,%
]{revtex4-1}
\usepackage{siunitx}
\usepackage{graphicx}% Include figure files
\usepackage{dcolumn}% Align table columns on decimal point
\usepackage{bm}% bold math
\begin{document}

\preprint{APS/123-QED}

\title{Generation of subcycle isolated attosecond pulses by pumping ionizing gating}% Force line breaks with \\
%\thanks{Footnote to title of article.}
\author{Zhaohui Wu}
\affiliation{Science and Technology on Plasma Physics Laboratory, Research Center of Laser Fusion, China Academy of Engineering Physics, Mianyang, Sichuan, China, 621900.}
\author{Hao Peng}
\affiliation{College of Physics and Optoelectronic Engineering, Shenzhen University, Shenzhen 518060, China}
\thanks{The first two authors contribute equally to this work}
\author{Xiaoming Zeng}
\author{Zhaoli Li}
\author{Zhimeng Zhang}
\affiliation{Science and Technology on Plasma Physics Laboratory, Research Center of Laser Fusion, China Academy of Engineering Physics, Mianyang, Sichuan, China, 621900.}
\author{Huabao Cao}
\author{Yuxi Fu}
\affiliation{Center for Attosecond Science and Technology, Xi’an Institute of Optics and Precision Mechanics, Chinese Academy of Sciences, Xi’an 710119, Shaanxi, China}
%\affiliation{University of Chinese Academy of Sciences, Beijing 100049, China}
\author{Xiaodong Wang}
\author{Xiao Wang}%
\author{Jie Mu}
\author{Yanlei Zuo}%
\email{zuoyanlei@tsinghua.org.cn}
\affiliation{Science and Technology on Plasma Physics Laboratory, Research Center of Laser Fusion, China Academy of Engineering Physics, Mianyang, Sichuan, China, 621900.}
\author{C. Riconda}
\affiliation{LULI, Sorbonne Université, CNRS, École Polytechnique, CEA, F-75005, Paris, France}
\author{S. Weber}
\affiliation{ELI Beamlines facility, Extreme Light Infrastructure ERIC, 25241 Dolni Brezany, Czech Republic }
\author{Jingqin Su}
\affiliation{Science and Technology on Plasma Physics Laboratory, Research Center of Laser Fusion, China Academy of Engineering Physics, Mianyang, Sichuan, China, 621900.}
\date{\today}% It is always \today, today,

\begin{abstract}
We present a novel approach named as pumping ionizing gating (PIG) for the generation of isolated attosecond pulses (IAPs). In this regime, a short laser is used to ionize a pre-existing gas grating, creating a fast-extending plasma grating(FEPG) having an ionization front propagating with the velocity of light. A low-intensity long counterpropagating pump pulse is then reflected by a very narrow region of the ionization front, only where the Bragg conditions for resonant reflection is satisfied. Consequently, the pump reflection is confined within a sub-cycle region called PIG, and forms a wide-band coherent IAP in combination with the frequency up-conversion effect due to the plasma gradient. This approach results in a new scheme to generate IAPs from long picosecond pump pulses. Three-dimensional (3D) simulations show that a 1.6-ps, 1-$\mu$m pump pulse can be used to generate a 330 as laser pulse with a peak intensity approximately 33 times that of the pump and a conversion efficiency of around 0.1$\%$.These results highlight the potential of the PIG method for generating IAPs with high conversion efficiency and peak intensity.
\end{abstract}

%Use showkeys class option if keyword
\maketitle

Ultrafast techniques have enabled the generation of laser pulses with durations on the attosecond scale, allowing for the direct observation of electron motion in various systems such as atoms, molecules, and solids. Isolated attosecond pulses (IAPs) are typically generated through gas high harmonics generation(GHHG) with gating techniques\cite{Francesca16,Michael14,Hentschel01,Sansone06,Goulielmakis04,Gaumnitz17,Witting12,Corkum94,Jie17,Chang04,Chang05,Sola06,Kun12,Chang07,Mashiko08,Mashiko12,Kyung13,Heyl14,Zhong16,Tzallas07,Ximao09,Skantzakis09,Mashiko13,Tzallas07}, which have spectral coverage from hard X-rays to extreme ultraviolet (EUV,$<120$ nm) and durations as short as a few tens of attoseconds. However, the pulse energies of these IAPs are limited to the nanojoule range due to the low efficiency of high harmonics generation. Alternative methods for generating IAPs include coherent field synthesis\cite{Huang11,Hassan16,Giulio20}, which has been demonstrated to produce pulses with durations of as short as 380 attoseconds and spectra covering the deep ultraviolet to near infrared range\cite{Hassan16}. It leads to longer pulse durations while allowing for higher pulse energies (tens of microjoules). Further increasing of the IAPs energy opens new possibilities for the nonlinear attosecond optics. However, it is limited by the energy of the few-cycle driving pulses.

 The key principle behind GHHG for generating IAPs lies in the use of gating techniques to confine the resonant generation zone to a narrow temporal region. However, this inherently results in low energy transfer efficiency since a significant portion of the driving pulse is not utilized. On the other hand, plasma has been shown to be an effective media for laser compression by exploiting the plasma waves\cite{Malkin991,Malkin00,Trines111,Ping04,Cheng05,JUN07,Turbunll18s,Wu20,Andreev06,Weber13,Lancia13,Lancia16,Marques19} or fast-extending plasma grating(FEPG)\cite{Zhaohui22,Zhaoli22}. This technique enables laser compression with high peak power and efficiency. However, the shortest duration of the output pulse is limited to several optical cycles.

\begin{figure}[htpb]
\centering
\includegraphics[width=3.2 in]{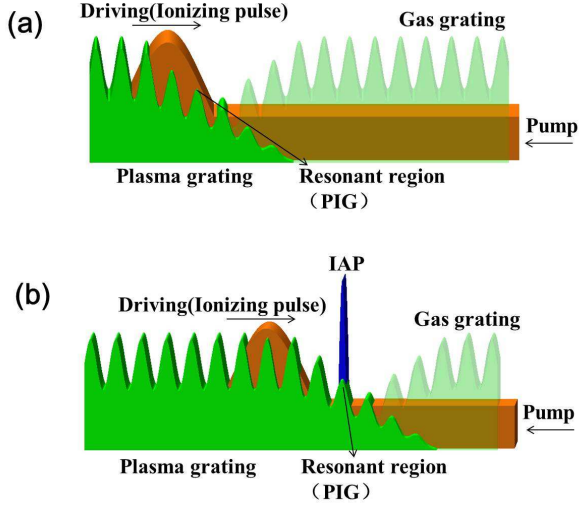}
\caption{(a) Illustration of the pumping ionizing gating(PIG) formed by the gradient FEPG. (b) Scheme of IAP generation due to pump reflection and frequency upconversion in the resonant region(PIG).
\label{fig:scheme}}
\end{figure}
In this letter, we propose a novel approach making joule-level IAPs possible by incorporating ionization gating techniques into the plasma compression using a gradient FEPG. The FEPG is formed by utilizing a short laser pulse to ionize a background gas grating, which can be created using established methods such as stimulated Brillouin scattering (SBS) \cite{Zhaohui22,Zhaoli22,SBS,Gorbunov83,yang20,Carr85,Hui91} or interference ionization \cite{chaojie21} in the gas medium. Following the ionization pulse, the plasma grating has a boundary extending at the light speed. In previous studies, the pump beam is resonantly reflected in the bulk plasma grating region to ensure pump depletion \cite{Zhaohui22,Zhaoli22}. However, in our proposed scheme, as illustrated in Fig.\ref{fig:scheme}(a), the resonant reflection region of the pump is confined to a narrow region at the ionization front by selecting a proper grating period. Within this ionization front, the gas grating is gradually ionized by the driving pulse, resulting in an increasing average plasma density $n_{e0}\rho(\xi)$, where $\xi=z-tv_g$, $v_g$ is the group velocity of the driving pulse, $\rho$ is the ionization rate, and ${n_e}_0$ is average electron density of the fully-ionized gas. For a grating with a period $\Lambda$, the Bragg condition for resonant reflection of the pump can only be fulfilled in a very narrow region around $\xi\approx \xi_0$, given by $k_0N_0\Lambda=\pi$, where $k_0=2\pi/\lambda_0$ and $\lambda_0$ are the pump wavevector and wavelength, and $N_0=(1+\sqrt{1-{n_e}_0\rho(\xi_0)/n_c})/2$ represents the refractive index at the resonant region  of the FEPG\cite{Zhaoli22}. Moreover, as the laser propagates through the gradient plasma, a frequency-upconversion effect occurs \cite{Kenan18,Kenan19,Howard19,Esarey91,Peng21}. Consequently, the reflected pump beam is confined and up-converted in the narrow resonant region, resulting in a wide-band coherent IAP, as shown Fig.\ref{fig:scheme}(b). Since the reflected pulse extracts energy from the entire long pump pulse, we call the approach pumping ionizing gating(PIG) which enables the IAP energy several orders of magnification higher than the present level.

To explain the physical model of PIG, we firstly consider the energy transfer from the pump to driving pulse by the FEPG, which is given by the coupled-waves equation in the variable of $\xi$\cite{Zhaohui22}:
\begin{eqnarray}
\begin{array}{rcl}
&&\frac{\partial E}{\partial t}=-i\frac{\omega^2_{pe}}{2\omega_0}E_0fe^{2i\omega_0\delta Nt},\\
&&\frac{\partial E_0}{\partial t}-2v_g\frac{\partial E_0}{\partial\xi}=-i\frac{\omega^2_{pe}}{2\omega_0}Efe^{-2i\omega_0\delta Nt},
\end{array}
\label{FEPG}
\end{eqnarray}
where $E$ and $E_0$ are the electric filed of driving pulse and pump pulse respectively, $\omega_{pe}=\sqrt{n_ee^2/m_e\varepsilon_0}$ is the plasma frequency, $n_e$ is the electron density, $e$ and $m_e$ are the electron charge and mass, $\varepsilon_0$ is the permittivity,$f\equiv\frac{\delta{n_e}_0}{{n_e}_{0}}$ is the normalized FEPG amplitude,  $\delta{n_e}_0$  is the amplitude  of the fully-ionized plasma grating, and $\omega_0$ is the pump frequency. 

The electron density $n_e$ in the gradient FEPG can be written as
\begin{eqnarray}
\begin{array}{rcl}
n_e=\left\{
\begin{array}{rcl}
{n_e}_0+\delta{n_e}_0sin(2\pi\xi/\Lambda),& & \xi<-L,\\
\rho[{n_e}_0+\delta{n_e}_0sin(2\pi\xi/\Lambda)], & & -L\leq\xi\leq0,\\
0, & & \xi>0,
\end{array}\right.
\end{array}
\label{ne}
\end{eqnarray}
where $L$ is the length of the ionization front. 

 As most regions of the gradient FEPG can not satisfy the resonant condition, there is a detuning factor of $e^{2i\omega_0\delta Nt}$ in the Eq.\ref{FEPG}, where $\delta N=(N-N_0)$, $N=(1+\sqrt{1-\rho(\xi){n_e}_0/n_c})/2$ is the refraction index of the pump in FEPG\cite{Zhaoli22}, $n_c=\omega^2_0m_e\varepsilon_0/{n_e}_0e^2$is the critical plasma density for the pump. One has $2k_0N_0\Lambda=2\pi$, therefore 
\begin{eqnarray}
\begin{array}{rcl}
\Lambda=\lambda_0/(1+\sqrt{1-\rho(\xi_0){n_e}_0/n_c}).
\end{array}
\label{Braggconditon}
\end{eqnarray}
 From Eq.\ref{FEPG} and Eq.\ref{Braggconditon} it can be seen that the growth rate of the driving pulse decreases exponentially at the position away from the resonant point, therefore a narrow gating is created, and it moves to the position a higher $\rho$ as $\Lambda$ increases. For $\rho$=1, Eq.\ref{Braggconditon} becomes the standard Bragg condition of FEPG\cite{Zhaoli22}.

 In a medium with a time-varying refractive index, the laser experiences a frequency shift while maintaining a constant wavevector in order to satisfy the dispersion relation \cite{Esarey91}. Theoretical models have been developed to describe this phenomenon in the ionizing plasma \cite{Kenan18,Kenan19,Howard19}.  In the case of the ionizing plasma grating, although the periodic structure also provides a time-varying refractive index, the integration over one period is close to zero. Consequently, the frequency upconversion effect resulting from the grating structure can be neglected, allowing us to extend the existing theoretical models to the gradient FEPG. To validate this assumption, we performed particle-in-cell (PIC) simulations using the same laser pulse to ionize the gas and the gas grating. The up-converted spectra were found to be nearly identical(see Fig.S1 in the supplement materials).

For a laser induced plasma, a roughly linear density gradient is formed in the pulse front. The up-converted light has an analytical solution of $\frac{\omega_0}{\omega}Ee^{i\omega t(\xi-\xi_0)}$ when $\omega_{pe}\ll\omega$, where $\omega=\omega_0(1+ct\omega^2_{pe}/L\omega_0^2)^{1/2}$ is the shifted laser frequency. However, the up-conversion process in PIG is more complex due to the reflected pump. As the reflected pump is mainly affected by the plasma gradient and the propagation time, it can be approximately divided into a series of independent infinitesimal pulses. For $dE$ at time $t$, the up-converted electric field at time $t+\delta t$ can be described by
\begin{eqnarray}
\begin{array}{rcl}
dE(t+\delta t)=\frac{\omega_0}{\omega(\delta t)} dE e^{i\omega(\delta t)(\xi-\xi_0)}.
\end{array}
\label{upconversion1}
\end{eqnarray}

 As the resonant region in the gradient FEPG is very narrow, the pump reflectivity is small and $E_0$ is approximate to a constant in Eq.\ref{FEPG}, so the driving pulse has an analytical  solution by combing Eq.\ref{FEPG} and Eq.\ref{upconversion1}.  
\begin{eqnarray}
\begin{array}{rcl}
E=\int^t_0i\frac{\omega^2_{pe}}{2\omega}E_0fe^{2i\omega_0\delta Nt+i\omega(\xi-\xi_0)}dt.
\end{array}
\label{IAP}
\end{eqnarray}
The detailed derivation of Eq.\ref{IAP} is provided in the supplement material.
 
 In order to further visualize the mechanism, PIC simulations were carried out using the code EPOCH\cite{Arber15}. A picosecond flat-top pulse with intensity $I_0=4\times10^{13}~\rm{W/cm^2}$ and wavelength $\lambda_0=1~\mu$m was used as the pump, while a femtosecond Gaussian pulse with intensity $10^{14}~\rm{W/cm^2}$  and an arbitrary wavelength served as the driving(ionizing) pulse. The background gas grating was set to hydrogen with a modulation period near  $\sim\lambda_0/2$. The ionization process was modeled using the Ammosov-Delone-Krainov tunnel ionization model\cite{Ammosov86}.

\begin{figure}[htpb]
\centering
\includegraphics[width=3 in]{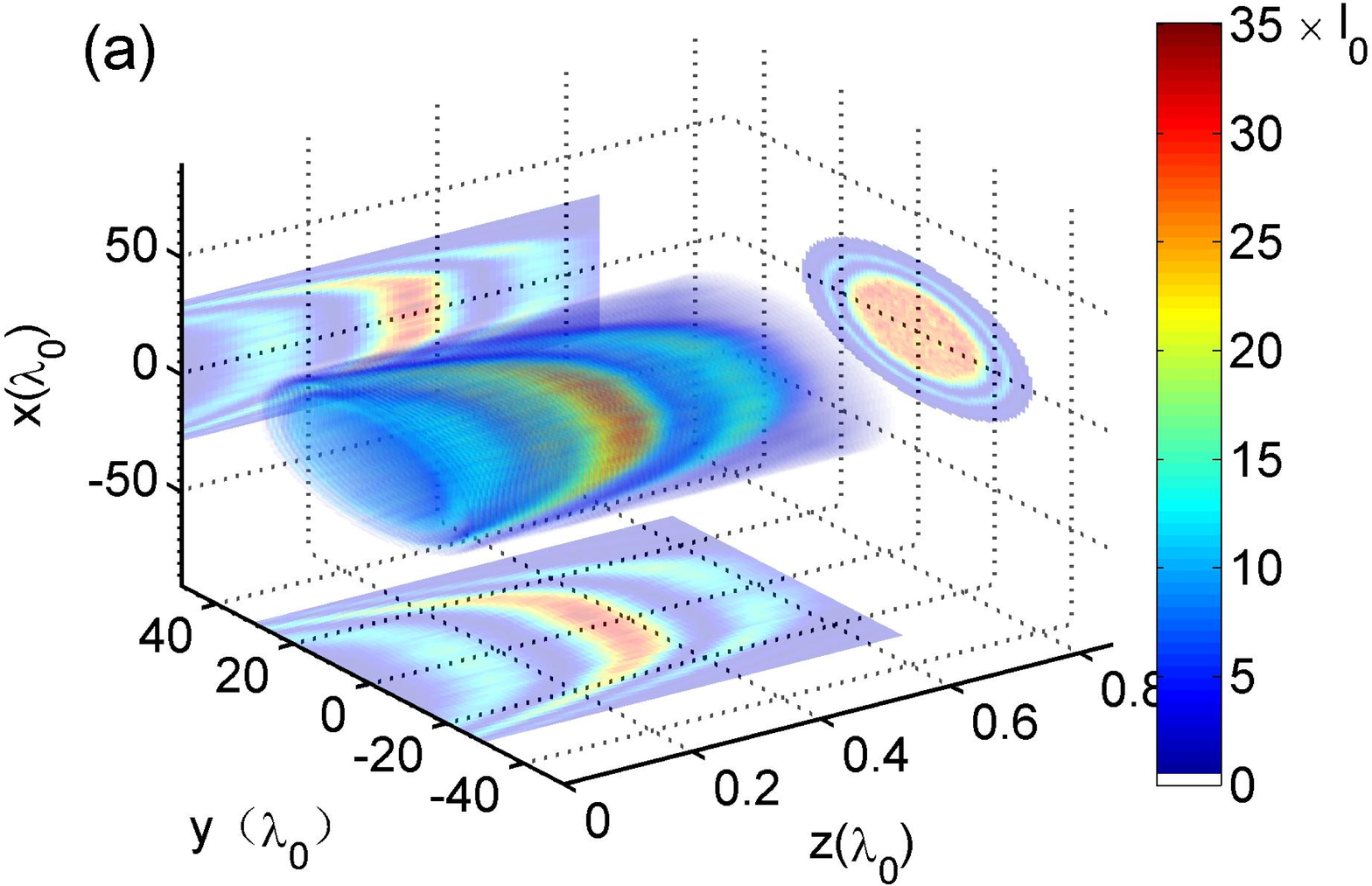}\\
\includegraphics[width=3 in]{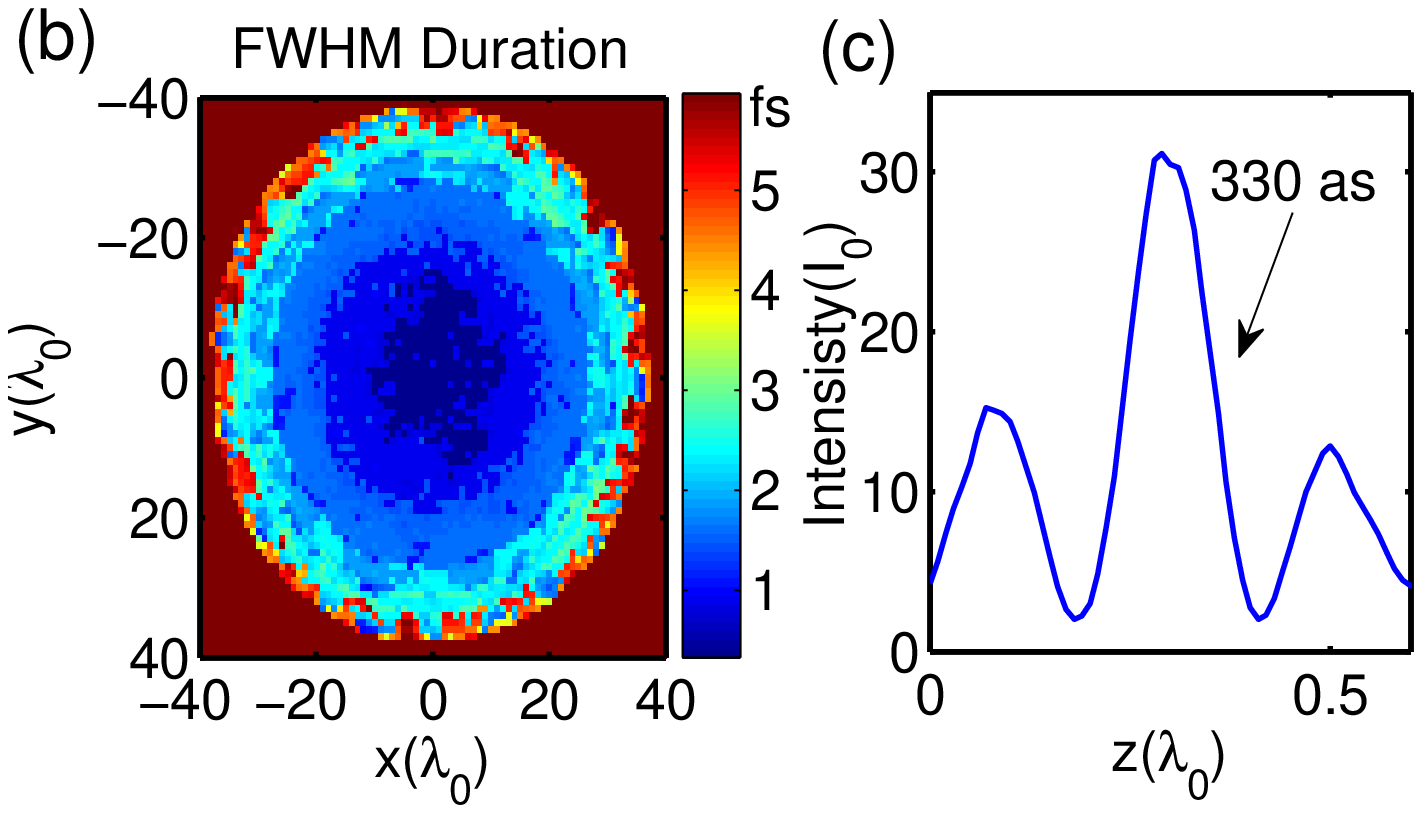}
\caption{ (a)Profiles of the 3D IAPs.(b)Distribution of FWHM pulse duration.(c)Temporal intensity of the IAP at the beam center.
\label{fig:3Das}}
\end{figure}

We firstly show 3D PIC simulation with high efficiency energy transfer and generation of intense IAPs. Parametric scans of the plasma densities and grating amplitudes allowed to identify the optimal result of IAPs(see the supplement materials), and the optimal parameters are given in the Tab.S1 in the supplement materials. In the simulation, a static window of $200\lambda_0\times200\lambda_0$ for the transverse x and y directions and $250\lambda_0$ for the longitudinal z direction was applied. Each cell had a size of $\lambda_0\times\lambda_0\times\lambda_0/100$ and contained 16 particles. Both the pump and driving pulse beams had a 6-order super Gaussian transverse distribution with a beam waist of 50 $\mu$m. The output driving pulse has a FWHM duration of 330 as, and a peak intensity of $\sim33I_0$, as shown in Fig.\ref{fig:3Das}(c). The total conversion efficiency from the pump to the IAP is about $0.1\%$, which is only 1/7 of the 1D case.  This is mainly due to the imperfect formation of IAPs at the beam edge where both the intensity of the input driving pulse and the pump decrease, as shown in Fig.\ref{fig:3Das}(b). The simulation results suggest that the efficiency could be enhanced by employing a larger beam with a more uniform intensity distribution.

\begin{figure}[htpb]
\centering
\includegraphics[width=3 in]{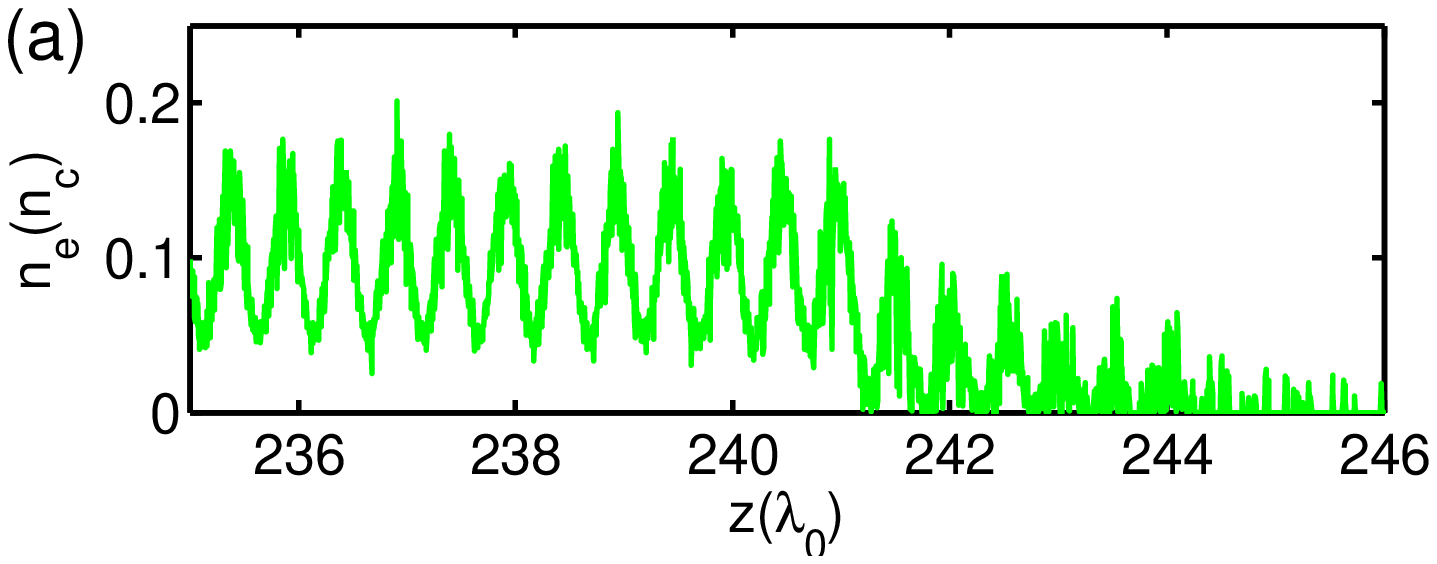}\\
\includegraphics[width=3 in]{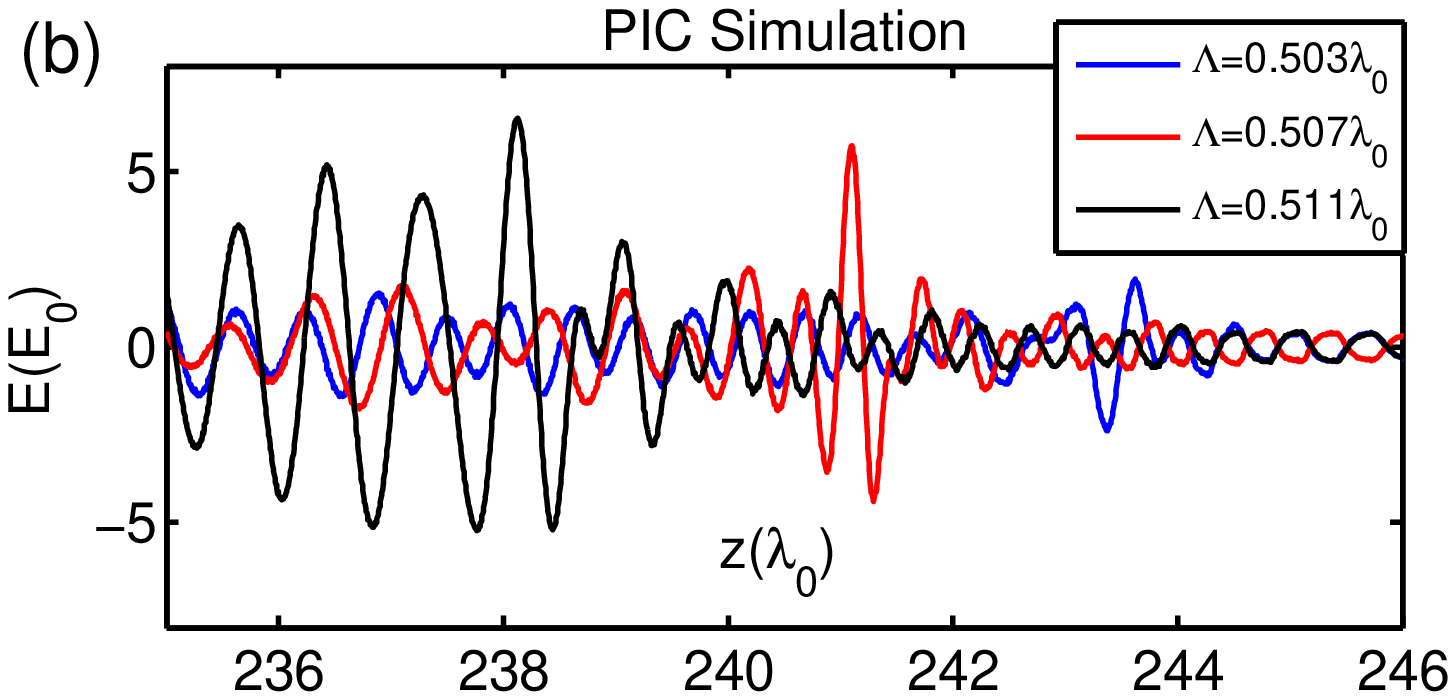}\\
\includegraphics[width=3 in]{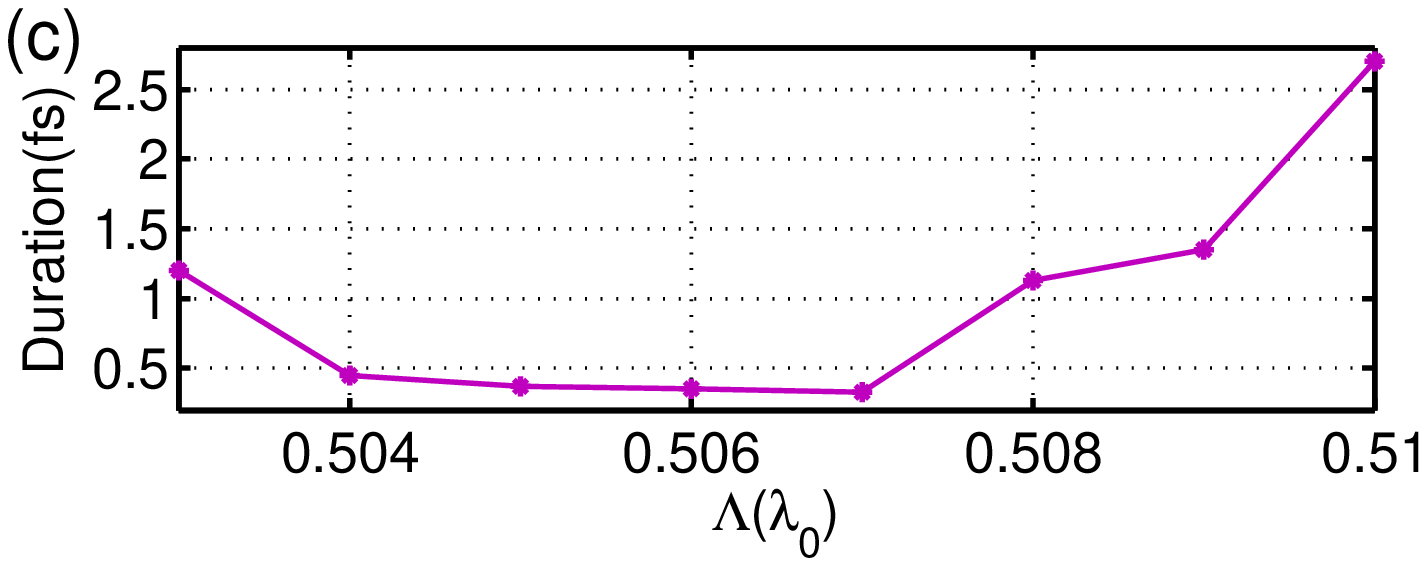}\\
\includegraphics[width=3 in]{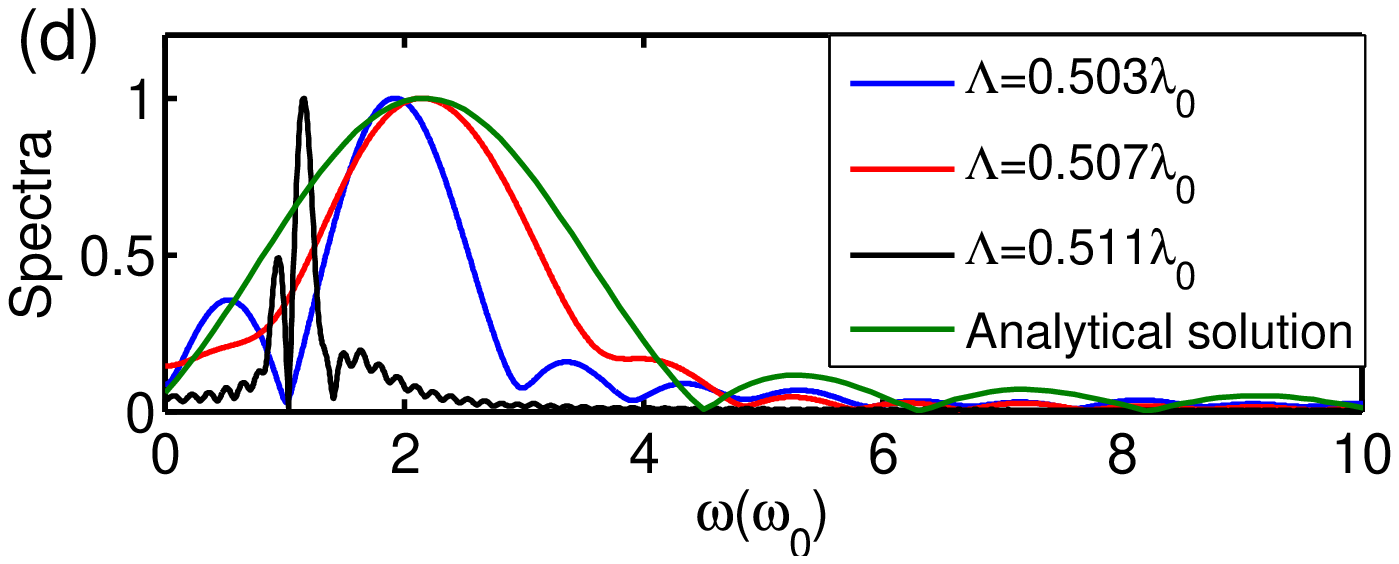}\\
\includegraphics[width=3 in]{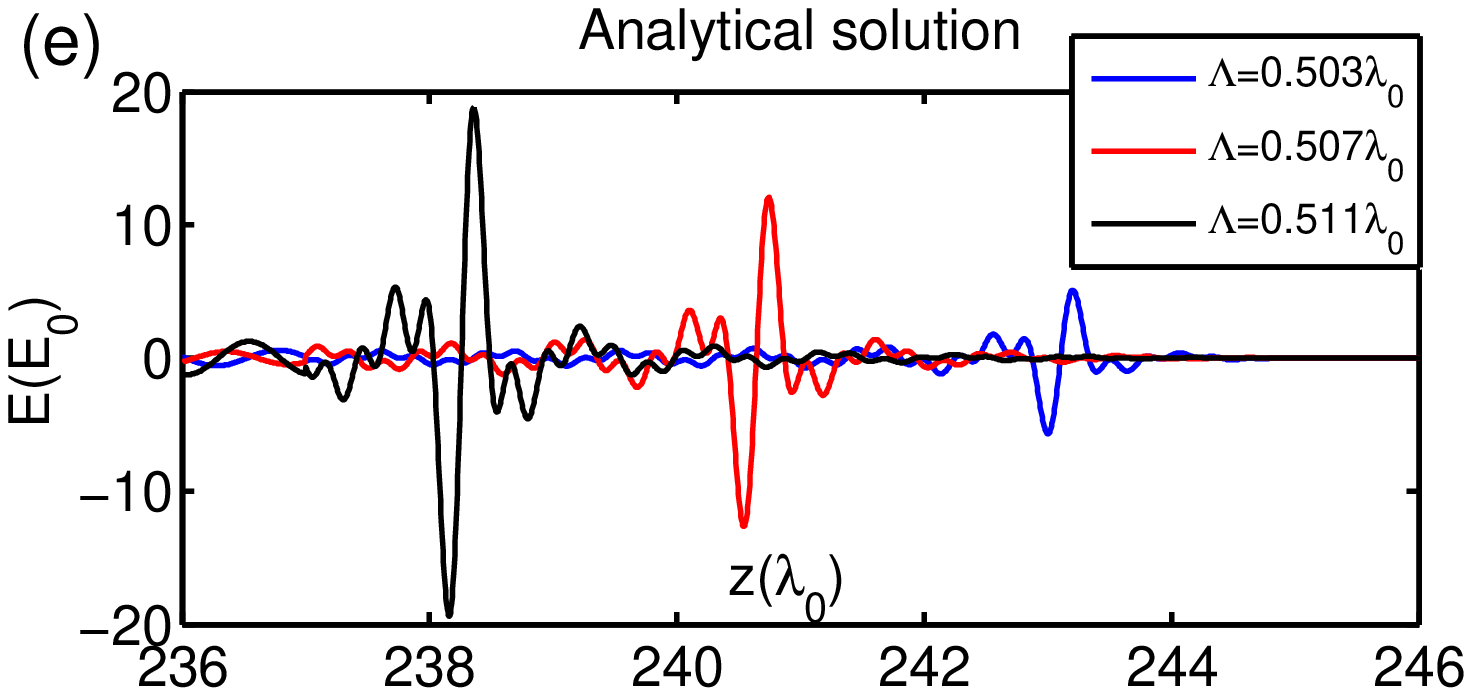}\\
\caption{ (a)Distribution of the electron density in the gradient FEPG. (b)Electric field of the driving-pulses at different $\Lambda$ obtained by the PIC simulation. (c)The shortest durations(FWHM) of the driving pulse with various $\Lambda$.(d)Spectra of the driving-pulses at different grating periods, and the analytical solution. (e)Wavefroms of the driving pulses at different $\Lambda$ obtained by the analytical solution.
\label{fig:1Dresult}}
\end{figure}

Varying the FEPG periods $\Lambda$, some typical features of PIG are observed as described below: Firstly, the output driving pulse shifts towards regions of higher average density as $\Lambda$ increases, with positions agreeing well with those of the resonant points predicted by Eq.\ref{Braggconditon}, as demonstrated in Fig.\ref{fig:1Dresult}(b) where several profiles of the driving pulses at different $\Lambda$ are gathered together. Secondly, the pulse amplitude rises with $\Lambda$, mainly due to a stronger reflection of the pump at resonant regions with higher average density, as depicted in Fig.\ref{fig:1Dresult}(b). Thirdly, the pulse duration firstly swiftly decreases to attosecond and then increases again. As $\Lambda$ increases, the PIG gradually forms, enabling the generation of IAPs within the range of $0.504\lambda_0\leq\Lambda\leq0.507\lambda_0$. However, when the resonant region is further shifted towards the fully ionized zone as $\Lambda$ increases, the range becomes wider due to a smaller density gradient. Consequently, the pulse duration exhibits a rebound effect after achieving its shortest value, as depicted in Fig.\ref{fig:1Dresult}(c). In experimental setups, the precise adjustment of $\Lambda$ can be achieved by varying the incident angle of the pump.

The frequency upconversion effect plays a significant role in the IAPs generation since it can provide a much wider spectrum for the output laser pulse. The PIC simulation shows that the central frequency of the IAP is shifted to around 2$\omega_0$[see Fig.\ref{fig:1Dresult}(d)], and the pulse duration changes to more than 100 fs with the frequency over $1.8\omega_0$ absent, indicating the up-converted light is the key ingredient for IAPs. Moreover, we utilized the analytical model given by Eq.\ref{IAP} to investigate the influence of the upcoversion effect while it cannot be explicitly removed in the PIC simulation. The analytical results with different $\Lambda$ using Eq.\ref{IAP} with $L=8\lambda_0$ is presented in Fig.\ref{fig:1Dresult}(d) and Fig.\ref{fig:1Dresult}(e). They exhibit good agreement with the PIC simulation in terms of waveform profiles and pulse positions, and the central frequency of the IAP is also shifted to around 2$\omega_0$. Note that the pulse spectra remains constant in the analytical model, whereas they vary in the PIC simulation with changing $\Lambda$. This discrepancy can be attributed to the assumption of a linear plasma gradient in the analytical model, whereas the PIC simulation incorporates a nonlinear gradient. Despite this difference, the analytical result can provide a qualitative investigation of the frequency upconversion effect. By disabling the upconversion effect in the analytical model, the pulse duration increases from hundreds of attosecond to over 1 fs, implying that it is indispensable for IPAs.

PIG is available in a wide parameter range as it can be formed whenever any region along the gradient FEPG satisfies the resonant condition. To confirm this, we summarize the simulation results with different average densities and modulation depths in Tab.\ref{tab:robustness}(more details are given by Fig.S2 in the supplemental material). The results show that IAPs can be obtained at a wide range  of average plasma densities(0.02$n_c$-0.5$n_c$) and modulation depths($60\%{n_e}_0$-$10\%{n_e}_0$). As shown in Tab.\ref{tab:robustness}, the required interaction length for IAP decrease with increasing average plasma density, indicating that higher plasma densities are preferred for higher conversion efficiency. However, the simulations reveal that pulse durations become unstable and peak intensities decrease when ${n_e}_0>0.1n_c$. Thus, ${n_e}_0=0.1n_c$ is identified as the optimal plasma density. Furthermore,  IAPs can still be generated at a modulation depth as small as $10\%{n_e}_0$, indicating that the creation of the required gas grating for IAPs can be significantly simplified.

\begin{table}[htpb]
\centering
\caption{\label{tab:robustness}Summary of 1D PIC simulation results with various average plasma density(${n_e}_0$) and modulation depths($\delta{n_e}_0/{n_e}_0$), where $L_i$ is the interaction length, $\tau$ is the FWHM duration of the driving pulse}
\begin{tabular}{ccccccccc}
\hline
\hline
 Case& I & II & III &IV &V&VI&VII\\
\hline
${n_e}_0(n_c)$ & 0.02&0.05 &0.1& 0.5 & 0.1 &0.1&0.1  \\
$\delta{n_e}_0/{n_e}_0$ & 0.5&0.5&0.5 & 0.5 & 0.1&0.2 &0.6 \\
$\Lambda$ & 0.5016&0.5035 &0.507& 0.535 & 0.506&0.506 &0.507  \\
$L_i(mm)$&0.9&0.44 &0.24&0.24 &0.26&0.24&0.22\\
 Peak $E(E_0)$ &4.7&5.4&5.75&4.4&2.6&4.4& 5.5 \\
 $\tau$(fs)&0.42&0.38&0.33&0.32& 0.42&0.4& 0.35\\
 efficiency($\%$) &0.15&0.39&0.7& 1.3& 0.14&0.48 & 0.72 \\
\hline
\hline
\end{tabular}
\end{table}

\begin{figure}[htpb]
\centering
\includegraphics[width=3.2 in]{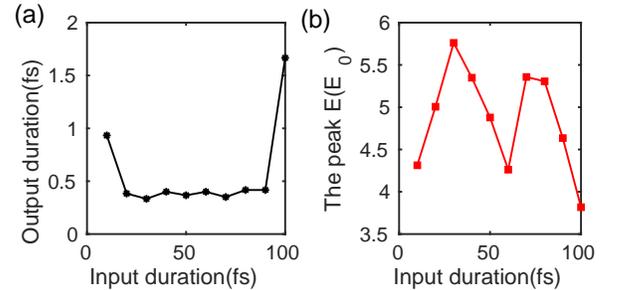}
\caption {Simulation results of the output FWHM duration $\tau$ (a) and the peak $E$(b) of the driving pulse with various input FWHM durations $\tau_{in}$.
\label{fig:duration}}
\end{figure}

A sufficiently short input driving pulse is required to form a narrow resonant region. However, the pulse cannot be excessively short, or the resonant region would be less than one period, rendering it inefficient for forming a Bragg grating. In the PIC simulation, IAPs are obtained with the input pulse durations $\tau_{in}$ ranging from 20 fs to 90 fs, as depicted in Fig.\ref{fig:duration}(a). The peak $E$ firstly increases and then drops as $\tau_{in}$ increases, with the maximum value of 5.75$E_0$ at $\tau_{in}=30$ fs. However, as shown in Eq.\ref{IAP} and Fig.\ref{fig:1Dresult}(b), the peak $E$ is also influenced by the grating period $\Lambda$ which slightly varies with $\tau_{in}$ to obtain optimal IAPs. Consequently, there is a dip of the peak $E$ at $\tau_{in}$ of 50-60 fs when $\Lambda$ is relative smaller, as shown in Fig.\ref{fig:duration}(b). In addition, a high contrast is not required, given that the peak intensity of the input driving pulse only slightly surpasses the ionization threshold of the background gas (approximately $10^{14}~\rm{W/cm^2}$). The requisite driving pulse can be readily provided by commercial Ti:sapphire CPA laser facilities.

In summary, a method for generating IAPs, referred to as the PIG mechanism, has been proposed. This approach involves the reflection of a picosecond pump pulse by a gradient FEPG where a sub-cycle gating is formed in the narrow resonant region. In corporation with the frequency up-conversion effect of the laser in the gradient plasma, it allows for the generation of sub-cycle IAPs. The feasibility of the approach has been demonstrated through PIC simulations, showing that IAPs can be generated across a wide range of average plasma densities (0.02$n_c$-0.5$n_c$) and modulation depths (60$\%{n_e}_0$-10$\%{n_e}_0$). The 3D simulation results obtains a 330-as IAP in a 240 $\mu$m-long, $0.1n_c$-density FEPG, with a conversion efficiency of $\sim0.1\%$. Using a 1-kJ, 1.6-ps, 1053-nm laser pulse produced by chirped pulse amplification\cite{Mourou85,Mourou12} as the pump, would allow to generate joule-level IAPs. 

This work was partly supported by National Key Program for S$\&$T Research and Development (Grant No. 2018YFA0404804), and the Science and Technology on Plasma Physics Laboratory Fund(Grant No. 6142A04220203 and 6142A04220204)

\bibliography{ref}% Produces the bibliography via BibTeX.

\providecommand{\noopsort}[1]{}\providecommand{\singleletter}[1]{#1}%
\begin{thebibliography}{56}%
\makeatletter
\providecommand \@ifxundefined [1]{%
 \@ifx{#1\undefined}
}%
\providecommand \@ifnum [1]{%
 \ifnum #1\expandafter \@firstoftwo
 \else \expandafter \@secondoftwo
 \fi
}%
\providecommand \@ifx [1]{%
 \ifx #1\expandafter \@firstoftwo
 \else \expandafter \@secondoftwo
 \fi
}%
\providecommand \natexlab [1]{#1}%
\providecommand \enquote  [1]{``#1''}%
\providecommand \bibnamefont  [1]{#1}%
\providecommand \bibfnamefont [1]{#1}%
\providecommand \citenamefont [1]{#1}%
\providecommand \href@noop [0]{\@secondoftwo}%
\providecommand \href [0]{\begingroup \@sanitize@url \@href}%
\providecommand \@href[1]{\@@startlink{#1}\@@href}%
\providecommand \@@href[1]{\endgroup#1\@@endlink}%
\providecommand \@sanitize@url [0]{\catcode `\\12\catcode `\$12\catcode
  `\&12\catcode `\#12\catcode `\^12\catcode `\_12\catcode `\%12\relax}%
\providecommand \@@startlink[1]{}%
\providecommand \@@endlink[0]{}%
\providecommand \url  [0]{\begingroup\@sanitize@url \@url }%
\providecommand \@url [1]{\endgroup\@href {#1}{\urlprefix }}%
\providecommand \urlprefix  [0]{URL }%
\providecommand \Eprint [0]{\href }%
\providecommand \doibase [0]{http://dx.doi.org/}%
\providecommand \selectlanguage [0]{\@gobble}%
\providecommand \bibinfo  [0]{\@secondoftwo}%
\providecommand \bibfield  [0]{\@secondoftwo}%
\providecommand \translation [1]{[#1]}%
\providecommand \BibitemOpen [0]{}%
\providecommand \bibitemStop [0]{}%
\providecommand \bibitemNoStop [0]{.\EOS\space}%
\providecommand \EOS [0]{\spacefactor3000\relax}%
\providecommand \BibitemShut  [1]{\csname bibitem#1\endcsname}%
\let\auto@bib@innerbib\@empty
%</preamble>
\bibitem [{\citenamefont {Calegari}\ \emph {et~al.}(2016)\citenamefont
  {Calegari}, \citenamefont {Sansone}, \citenamefont {Stagira}, \citenamefont
  {Vozzi},\ and\ \citenamefont {Nisoli}}]{Francesca16}%
  \BibitemOpen
  \bibfield  {author} {\bibinfo {author} {\bibfnamefont {F.}~\bibnamefont
  {Calegari}}, \bibinfo {author} {\bibfnamefont {G.}~\bibnamefont {Sansone}},
  \bibinfo {author} {\bibfnamefont {S.}~\bibnamefont {Stagira}}, \bibinfo
  {author} {\bibfnamefont {C.}~\bibnamefont {Vozzi}}, \ and\ \bibinfo {author}
  {\bibfnamefont {M.}~\bibnamefont {Nisoli}},\ }\href@noop {} {\bibfield
  {journal} {\bibinfo  {journal} {Journal of Physics B: Atomic, Molecular and
  Optical Physics}\ }\textbf {\bibinfo {volume} {49}},\ \bibinfo {pages}
  {062001} (\bibinfo {year} {2016})}\BibitemShut {NoStop}%
\bibitem [{\citenamefont {Chini}, \citenamefont {Zhao},\ and\ \citenamefont
  {Chang}(2014)}]{Michael14}%
  \BibitemOpen
  \bibfield  {author} {\bibinfo {author} {\bibfnamefont {M.}~\bibnamefont
  {Chini}}, \bibinfo {author} {\bibfnamefont {K.}~\bibnamefont {Zhao}}, \ and\
  \bibinfo {author} {\bibfnamefont {Z.}~\bibnamefont {Chang}},\ }\href@noop {}
  {\bibfield  {journal} {\bibinfo  {journal} {Nat. Photonics}\ }\textbf
  {\bibinfo {volume} {8}},\ \bibinfo {pages} {178} (\bibinfo {year}
  {2014})}\BibitemShut {NoStop}%
\bibitem [{\citenamefont {Hentschel}\ \emph {et~al.}(2001)\citenamefont
  {Hentschel}, \citenamefont {Kienberger}, \citenamefont {Spielmann},
  \citenamefont {Reider}, \citenamefont {Milosevic}, \citenamefont {Brabec},
  \citenamefont {Corkum}, \citenamefont {Heinzmann}, \citenamefont {Drescher},\
  and\ \citenamefont {Krausz}}]{Hentschel01}%
  \BibitemOpen
  \bibfield  {author} {\bibinfo {author} {\bibfnamefont {M.}~\bibnamefont
  {Hentschel}}, \bibinfo {author} {\bibfnamefont {R.}~\bibnamefont
  {Kienberger}}, \bibinfo {author} {\bibfnamefont {C.}~\bibnamefont
  {Spielmann}}, \bibinfo {author} {\bibfnamefont {G.~A.}\ \bibnamefont
  {Reider}}, \bibinfo {author} {\bibfnamefont {N.}~\bibnamefont {Milosevic}},
  \bibinfo {author} {\bibfnamefont {T.}~\bibnamefont {Brabec}}, \bibinfo
  {author} {\bibfnamefont {P.}~\bibnamefont {Corkum}}, \bibinfo {author}
  {\bibfnamefont {U.}~\bibnamefont {Heinzmann}}, \bibinfo {author}
  {\bibfnamefont {M.}~\bibnamefont {Drescher}}, \ and\ \bibinfo {author}
  {\bibfnamefont {F.}~\bibnamefont {Krausz}},\ }\href@noop {} {\bibfield
  {journal} {\bibinfo  {journal} {Nature}\ }\textbf {\bibinfo {volume} {414}},\
  \bibinfo {pages} {508} (\bibinfo {year} {2001})}\BibitemShut {NoStop}%
\bibitem [{\citenamefont {Sansone}\ \emph {et~al.}(2006)\citenamefont
  {Sansone}, \citenamefont {Benedetti}, \citenamefont {Calegari}, \citenamefont
  {Vozzi}, \citenamefont {Avaldi}, \citenamefont {Flammini}, \citenamefont
  {Poletto}, \citenamefont {Villoresi}, \citenamefont {Altucci}, \citenamefont
  {Velotta}, \citenamefont {Stagira}, \citenamefont {Silvestri},\ and\
  \citenamefont {Nisoli}}]{Sansone06}%
  \BibitemOpen
  \bibfield  {author} {\bibinfo {author} {\bibfnamefont {G.}~\bibnamefont
  {Sansone}}, \bibinfo {author} {\bibfnamefont {E.}~\bibnamefont {Benedetti}},
  \bibinfo {author} {\bibfnamefont {F.}~\bibnamefont {Calegari}}, \bibinfo
  {author} {\bibfnamefont {C.}~\bibnamefont {Vozzi}}, \bibinfo {author}
  {\bibfnamefont {L.}~\bibnamefont {Avaldi}}, \bibinfo {author} {\bibfnamefont
  {R.}~\bibnamefont {Flammini}}, \bibinfo {author} {\bibfnamefont
  {L.}~\bibnamefont {Poletto}}, \bibinfo {author} {\bibfnamefont
  {P.}~\bibnamefont {Villoresi}}, \bibinfo {author} {\bibfnamefont
  {C.}~\bibnamefont {Altucci}}, \bibinfo {author} {\bibfnamefont
  {R.}~\bibnamefont {Velotta}}, \bibinfo {author} {\bibfnamefont
  {S.}~\bibnamefont {Stagira}}, \bibinfo {author} {\bibfnamefont {S.~D.}\
  \bibnamefont {Silvestri}}, \ and\ \bibinfo {author} {\bibfnamefont
  {M.}~\bibnamefont {Nisoli}},\ }\href@noop {} {\bibfield  {journal} {\bibinfo
  {journal} {Science}\ }\textbf {\bibinfo {volume} {314}},\ \bibinfo {pages}
  {443} (\bibinfo {year} {2006})}\BibitemShut {NoStop}%
\bibitem [{\citenamefont {Goulielmakis}\ \emph {et~al.}(2004)\citenamefont
  {Goulielmakis}, \citenamefont {Schultze}, \citenamefont {Hofstetter},
  \citenamefont {Yakovlev}, \citenamefont {Gagnon}, \citenamefont {Uiberacker},
  \citenamefont {Aquila}, \citenamefont {Gullikson}, \citenamefont {Attwood},
  \citenamefont {Kienberger}, \citenamefont {Krausz},\ and\ \citenamefont
  {Kleineberg}}]{Goulielmakis04}%
  \BibitemOpen
  \bibfield  {author} {\bibinfo {author} {\bibfnamefont {E.}~\bibnamefont
  {Goulielmakis}}, \bibinfo {author} {\bibfnamefont {M.}~\bibnamefont
  {Schultze}}, \bibinfo {author} {\bibfnamefont {M.}~\bibnamefont
  {Hofstetter}}, \bibinfo {author} {\bibfnamefont {V.~S.}\ \bibnamefont
  {Yakovlev}}, \bibinfo {author} {\bibfnamefont {J.}~\bibnamefont {Gagnon}},
  \bibinfo {author} {\bibfnamefont {M.}~\bibnamefont {Uiberacker}}, \bibinfo
  {author} {\bibfnamefont {A.~L.}\ \bibnamefont {Aquila}}, \bibinfo {author}
  {\bibfnamefont {E.~M.}\ \bibnamefont {Gullikson}}, \bibinfo {author}
  {\bibfnamefont {D.~T.}\ \bibnamefont {Attwood}}, \bibinfo {author}
  {\bibfnamefont {R.}~\bibnamefont {Kienberger}}, \bibinfo {author}
  {\bibfnamefont {F.}~\bibnamefont {Krausz}}, \ and\ \bibinfo {author}
  {\bibfnamefont {U.}~\bibnamefont {Kleineberg}},\ }\href@noop {} {\bibfield
  {journal} {\bibinfo  {journal} {Nature}\ }\textbf {\bibinfo {volume} {427}},\
  \bibinfo {pages} {817} (\bibinfo {year} {2004})}\BibitemShut {NoStop}%
\bibitem [{\citenamefont {Gaumnitz}\ \emph {et~al.}(2017)\citenamefont
  {Gaumnitz}, \citenamefont {Jain}, \citenamefont {Pertot}, \citenamefont
  {Huppert}, \citenamefont {Jordan}, \citenamefont {Ardana-Lamas},\ and\
  \citenamefont {Worner}}]{Gaumnitz17}%
  \BibitemOpen
  \bibfield  {author} {\bibinfo {author} {\bibfnamefont {T.}~\bibnamefont
  {Gaumnitz}}, \bibinfo {author} {\bibfnamefont {A.}~\bibnamefont {Jain}},
  \bibinfo {author} {\bibfnamefont {Y.}~\bibnamefont {Pertot}}, \bibinfo
  {author} {\bibfnamefont {M.}~\bibnamefont {Huppert}}, \bibinfo {author}
  {\bibfnamefont {I.}~\bibnamefont {Jordan}}, \bibinfo {author} {\bibfnamefont
  {F.}~\bibnamefont {Ardana-Lamas}}, \ and\ \bibinfo {author} {\bibfnamefont
  {H.}~\bibnamefont {Worner}},\ }\href@noop {} {\bibfield  {journal} {\bibinfo
  {journal} {Opt. Express}\ }\textbf {\bibinfo {volume} {25}},\ \bibinfo
  {pages} {27506} (\bibinfo {year} {2017})}\BibitemShut {NoStop}%
\bibitem [{\citenamefont {Witting}\ \emph {et~al.}(2012)\citenamefont
  {Witting}, \citenamefont {Frank}, \citenamefont {Okell}, \citenamefont
  {Arrell}, \citenamefont {Marangos},\ and\ \citenamefont {Tisch}}]{Witting12}%
  \BibitemOpen
  \bibfield  {author} {\bibinfo {author} {\bibfnamefont {T.}~\bibnamefont
  {Witting}}, \bibinfo {author} {\bibfnamefont {F.}~\bibnamefont {Frank}},
  \bibinfo {author} {\bibfnamefont {W.~A.}\ \bibnamefont {Okell}}, \bibinfo
  {author} {\bibfnamefont {C.~A.}\ \bibnamefont {Arrell}}, \bibinfo {author}
  {\bibfnamefont {J.~P.}\ \bibnamefont {Marangos}}, \ and\ \bibinfo {author}
  {\bibfnamefont {J.~W.~G.}\ \bibnamefont {Tisch}},\ }\href@noop {} {\bibfield
  {journal} {\bibinfo  {journal} {Journal of Physics B: Atomic, Molecular and
  Optical Physics}\ }\textbf {\bibinfo {volume} {45}},\ \bibinfo {pages}
  {074014} (\bibinfo {year} {2012})}\BibitemShut {NoStop}%
\bibitem [{\citenamefont {Corkum}, \citenamefont {Burnett},\ and\ \citenamefont
  {Ivanov}(1994)}]{Corkum94}%
  \BibitemOpen
  \bibfield  {author} {\bibinfo {author} {\bibfnamefont {P.~B.}\ \bibnamefont
  {Corkum}}, \bibinfo {author} {\bibfnamefont {N.~H.}\ \bibnamefont {Burnett}},
  \ and\ \bibinfo {author} {\bibfnamefont {M.~Y.}\ \bibnamefont {Ivanov}},\
  }\href@noop {} {\bibfield  {journal} {\bibinfo  {journal} {Opt. Lett.}\
  }\textbf {\bibinfo {volume} {19}},\ \bibinfo {pages} {1870} (\bibinfo {year}
  {1994})}\BibitemShut {NoStop}%
\bibitem [{\citenamefont {Li}\ \emph {et~al.}(2017)\citenamefont {Li},
  \citenamefont {Ren}, \citenamefont {Yin}, \citenamefont {Zhao}, \citenamefont
  {Chew}, \citenamefont {Cheng}, \citenamefont {Cunningham}, \citenamefont
  {Wang}, \citenamefont {Hu}, \citenamefont {Wu}, \citenamefont {Chini},\ and\
  \citenamefont {Chang}}]{Jie17}%
  \BibitemOpen
  \bibfield  {author} {\bibinfo {author} {\bibfnamefont {J.}~\bibnamefont
  {Li}}, \bibinfo {author} {\bibfnamefont {X.}~\bibnamefont {Ren}}, \bibinfo
  {author} {\bibfnamefont {Y.}~\bibnamefont {Yin}}, \bibinfo {author}
  {\bibfnamefont {K.}~\bibnamefont {Zhao}}, \bibinfo {author} {\bibfnamefont
  {A.}~\bibnamefont {Chew}}, \bibinfo {author} {\bibfnamefont {Y.}~\bibnamefont
  {Cheng}}, \bibinfo {author} {\bibfnamefont {E.}~\bibnamefont {Cunningham}},
  \bibinfo {author} {\bibfnamefont {Y.}~\bibnamefont {Wang}}, \bibinfo {author}
  {\bibfnamefont {S.}~\bibnamefont {Hu}}, \bibinfo {author} {\bibfnamefont
  {Y.}~\bibnamefont {Wu}}, \bibinfo {author} {\bibfnamefont {M.}~\bibnamefont
  {Chini}}, \ and\ \bibinfo {author} {\bibfnamefont {Z.}~\bibnamefont
  {Chang}},\ }\href@noop {} {\bibfield  {journal} {\bibinfo  {journal} {Nat.
  Commun.}\ }\textbf {\bibinfo {volume} {8}},\ \bibinfo {pages} {186} (\bibinfo
  {year} {2017})}\BibitemShut {NoStop}%
\bibitem [{\citenamefont {Z}(2004)}]{Chang04}%
  \BibitemOpen
  \bibfield  {author} {\bibinfo {author} {\bibfnamefont {C.}~\bibnamefont
  {Z}},\ }\href@noop {} {\bibfield  {journal} {\bibinfo  {journal} {Phys. Rev.
  A}\ }\textbf {\bibinfo {volume} {70}},\ \bibinfo {pages} {043802} (\bibinfo
  {year} {2004})}\BibitemShut {NoStop}%
\bibitem [{\citenamefont {Z}(2005)}]{Chang05}%
  \BibitemOpen
  \bibfield  {author} {\bibinfo {author} {\bibfnamefont {C.}~\bibnamefont
  {Z}},\ }\href@noop {} {\bibfield  {journal} {\bibinfo  {journal} {Phys. Rev.
  A}\ }\textbf {\bibinfo {volume} {71}},\ \bibinfo {pages} {023813} (\bibinfo
  {year} {2005})}\BibitemShut {NoStop}%
\bibitem [{\citenamefont {Sola}\ \emph {et~al.}(2006)\citenamefont {Sola},
  \citenamefont {M\'{e}vel}, \citenamefont {Elouga}, \citenamefont {Constant},
  \citenamefont {Strelkov}, \citenamefont {Poletto}, \citenamefont {Villoresi},
  \citenamefont {Benedetti}, \citenamefont {Caumes}, \citenamefont {Stagira},
  \citenamefont {Vozzi}, \citenamefont {Sansone},\ and\ \citenamefont
  {Nisoli}}]{Sola06}%
  \BibitemOpen
  \bibfield  {author} {\bibinfo {author} {\bibfnamefont {I.~J.}\ \bibnamefont
  {Sola}}, \bibinfo {author} {\bibfnamefont {E.}~\bibnamefont {M\'{e}vel}},
  \bibinfo {author} {\bibfnamefont {L.}~\bibnamefont {Elouga}}, \bibinfo
  {author} {\bibfnamefont {E.}~\bibnamefont {Constant}}, \bibinfo {author}
  {\bibfnamefont {V.}~\bibnamefont {Strelkov}}, \bibinfo {author}
  {\bibfnamefont {L.}~\bibnamefont {Poletto}}, \bibinfo {author} {\bibfnamefont
  {P.}~\bibnamefont {Villoresi}}, \bibinfo {author} {\bibfnamefont
  {E.}~\bibnamefont {Benedetti}}, \bibinfo {author} {\bibfnamefont {J.-P.}\
  \bibnamefont {Caumes}}, \bibinfo {author} {\bibfnamefont {S.}~\bibnamefont
  {Stagira}}, \bibinfo {author} {\bibfnamefont {C.}~\bibnamefont {Vozzi}},
  \bibinfo {author} {\bibfnamefont {G.}~\bibnamefont {Sansone}}, \ and\
  \bibinfo {author} {\bibfnamefont {M.}~\bibnamefont {Nisoli}},\ }\href@noop {}
  {\bibfield  {journal} {\bibinfo  {journal} {Nat. Phys.}\ }\textbf {\bibinfo
  {volume} {9}},\ \bibinfo {pages} {319} (\bibinfo {year} {2006})}\BibitemShut
  {NoStop}%
\bibitem [{\citenamefont {Zhao}\ \emph {et~al.}(2012)\citenamefont {Zhao},
  \citenamefont {Zhang}, \citenamefont {Chini}, \citenamefont {Wu},
  \citenamefont {Wang},\ and\ \citenamefont {Chang}}]{Kun12}%
  \BibitemOpen
  \bibfield  {author} {\bibinfo {author} {\bibfnamefont {K.}~\bibnamefont
  {Zhao}}, \bibinfo {author} {\bibfnamefont {Q.}~\bibnamefont {Zhang}},
  \bibinfo {author} {\bibfnamefont {M.}~\bibnamefont {Chini}}, \bibinfo
  {author} {\bibfnamefont {Y.}~\bibnamefont {Wu}}, \bibinfo {author}
  {\bibfnamefont {X.}~\bibnamefont {Wang}}, \ and\ \bibinfo {author}
  {\bibfnamefont {Z.}~\bibnamefont {Chang}},\ }\href@noop {} {\bibfield
  {journal} {\bibinfo  {journal} {Opt. Lett.}\ }\textbf {\bibinfo {volume}
  {37}},\ \bibinfo {pages} {3891} (\bibinfo {year} {2012})}\BibitemShut
  {NoStop}%
\bibitem [{\citenamefont {Chang}(2007)}]{Chang07}%
  \BibitemOpen
  \bibfield  {author} {\bibinfo {author} {\bibfnamefont {Z.}~\bibnamefont
  {Chang}},\ }\href@noop {} {\bibfield  {journal} {\bibinfo  {journal} {Phys.
  Rev. A.}\ }\textbf {\bibinfo {volume} {76}},\ \bibinfo {pages} {051403(R)}
  (\bibinfo {year} {2007})}\BibitemShut {NoStop}%
\bibitem [{\citenamefont {Mashiko}\ \emph {et~al.}(2008)\citenamefont
  {Mashiko}, \citenamefont {Gilbertson}, \citenamefont {Li}, \citenamefont
  {Khan}, \citenamefont {Shakya}, \citenamefont {Moon},\ and\ \citenamefont
  {Chang}}]{Mashiko08}%
  \BibitemOpen
  \bibfield  {author} {\bibinfo {author} {\bibfnamefont {H.}~\bibnamefont
  {Mashiko}}, \bibinfo {author} {\bibfnamefont {S.}~\bibnamefont {Gilbertson}},
  \bibinfo {author} {\bibfnamefont {C.}~\bibnamefont {Li}}, \bibinfo {author}
  {\bibfnamefont {S.~D.}\ \bibnamefont {Khan}}, \bibinfo {author}
  {\bibfnamefont {M.~M.}\ \bibnamefont {Shakya}}, \bibinfo {author}
  {\bibfnamefont {E.}~\bibnamefont {Moon}}, \ and\ \bibinfo {author}
  {\bibfnamefont {Z.}~\bibnamefont {Chang}},\ }\href@noop {} {\bibfield
  {journal} {\bibinfo  {journal} {Phys Rev Lett}\ }\textbf {\bibinfo {volume}
  {100}},\ \bibinfo {pages} {103906} (\bibinfo {year} {2008})}\BibitemShut
  {NoStop}%
\bibitem [{\citenamefont {Vincenti}\ and\ \citenamefont
  {Qu\'{e}r\'{e}}(2012)}]{Mashiko12}%
  \BibitemOpen
  \bibfield  {author} {\bibinfo {author} {\bibfnamefont {H.}~\bibnamefont
  {Vincenti}}\ and\ \bibinfo {author} {\bibfnamefont {F.}~\bibnamefont
  {Qu\'{e}r\'{e}}},\ }\href@noop {} {\bibfield  {journal} {\bibinfo  {journal}
  {Phys Rev Lett}\ }\textbf {\bibinfo {volume} {108}},\ \bibinfo {pages}
  {113904} (\bibinfo {year} {2012})}\BibitemShut {NoStop}%
\bibitem [{\citenamefont {Kim}\ \emph {et~al.}(2013)\citenamefont {Kim},
  \citenamefont {Zhang}, \citenamefont {Ruchon}, \citenamefont {Hergott},
  \citenamefont {Auguste}, \citenamefont {Villeneuve}, \citenamefont {Corkum},\
  and\ \citenamefont {Qu\'{o}r\'{o}}}]{Kyung13}%
  \BibitemOpen
  \bibfield  {author} {\bibinfo {author} {\bibfnamefont {K.~T.}\ \bibnamefont
  {Kim}}, \bibinfo {author} {\bibfnamefont {C.}~\bibnamefont {Zhang}}, \bibinfo
  {author} {\bibfnamefont {T.}~\bibnamefont {Ruchon}}, \bibinfo {author}
  {\bibfnamefont {J.-F.}\ \bibnamefont {Hergott}}, \bibinfo {author}
  {\bibfnamefont {T.}~\bibnamefont {Auguste}}, \bibinfo {author} {\bibfnamefont
  {D.~M.}\ \bibnamefont {Villeneuve}}, \bibinfo {author} {\bibfnamefont
  {P.~B.}\ \bibnamefont {Corkum}}, \ and\ \bibinfo {author} {\bibfnamefont
  {F.}~\bibnamefont {Qu\'{o}r\'{o}}},\ }\href@noop {} {\bibfield  {journal}
  {\bibinfo  {journal} {Nat. Photonics}\ }\textbf {\bibinfo {volume} {7}},\
  \bibinfo {pages} {651} (\bibinfo {year} {2013})}\BibitemShut {NoStop}%
\bibitem [{\citenamefont {Heyl}\ \emph {et~al.}(2014)\citenamefont {Heyl},
  \citenamefont {Bengtsson}, \citenamefont {Carlstro\"{o}m}, \citenamefont
  {Mauritsson}, \citenamefont {Arnold},\ and\ \citenamefont
  {L\`{H}uillier;}}]{Heyl14}%
  \BibitemOpen
  \bibfield  {author} {\bibinfo {author} {\bibfnamefont {C.~M.}\ \bibnamefont
  {Heyl}}, \bibinfo {author} {\bibfnamefont {S.~N.}\ \bibnamefont {Bengtsson}},
  \bibinfo {author} {\bibfnamefont {S.}~\bibnamefont {Carlstro\"{o}m}},
  \bibinfo {author} {\bibfnamefont {J.}~\bibnamefont {Mauritsson}}, \bibinfo
  {author} {\bibfnamefont {C.~L.}\ \bibnamefont {Arnold}}, \ and\ \bibinfo
  {author} {\bibfnamefont {A.}~\bibnamefont {L\`{H}uillier;}},\ }\href@noop {}
  {\bibfield  {journal} {\bibinfo  {journal} {New J Phys}\ }\textbf {\bibinfo
  {volume} {16}},\ \bibinfo {pages} {052001} (\bibinfo {year}
  {2014})}\BibitemShut {NoStop}%
\bibitem [{\citenamefont {Zhong}\ \emph {et~al.}(2016)\citenamefont {Zhong},
  \citenamefont {He}, \citenamefont {Jiang}, \citenamefont {Teng},
  \citenamefont {He}, \citenamefont {Liu}, \citenamefont {Zhao},\ and\
  \citenamefont {Wei}}]{Zhong16}%
  \BibitemOpen
  \bibfield  {author} {\bibinfo {author} {\bibfnamefont {S.}~\bibnamefont
  {Zhong}}, \bibinfo {author} {\bibfnamefont {X.}~\bibnamefont {He}}, \bibinfo
  {author} {\bibfnamefont {Y.}~\bibnamefont {Jiang}}, \bibinfo {author}
  {\bibfnamefont {H.}~\bibnamefont {Teng}}, \bibinfo {author} {\bibfnamefont
  {P.}~\bibnamefont {He}}, \bibinfo {author} {\bibfnamefont {Y.}~\bibnamefont
  {Liu}}, \bibinfo {author} {\bibfnamefont {K.}~\bibnamefont {Zhao}}, \ and\
  \bibinfo {author} {\bibfnamefont {Z.}~\bibnamefont {Wei}},\ }\href@noop {}
  {\bibfield  {journal} {\bibinfo  {journal} {Phys. Rev. A}\ }\textbf {\bibinfo
  {volume} {93}},\ \bibinfo {pages} {033854} (\bibinfo {year}
  {2016})}\BibitemShut {NoStop}%
\bibitem [{\citenamefont {Tzallas}\ \emph {et~al.}(2007)\citenamefont
  {Tzallas}, \citenamefont {Skantzakis}, \citenamefont {Kalpouzos},
  \citenamefont {Benis}, \citenamefont {Tsakiris},\ and\ \citenamefont
  {Charalambidis}}]{Tzallas07}%
  \BibitemOpen
  \bibfield  {author} {\bibinfo {author} {\bibfnamefont {P.}~\bibnamefont
  {Tzallas}}, \bibinfo {author} {\bibfnamefont {E.}~\bibnamefont {Skantzakis}},
  \bibinfo {author} {\bibfnamefont {C.}~\bibnamefont {Kalpouzos}}, \bibinfo
  {author} {\bibfnamefont {E.~P.}\ \bibnamefont {Benis}}, \bibinfo {author}
  {\bibfnamefont {G.~D.}\ \bibnamefont {Tsakiris}}, \ and\ \bibinfo {author}
  {\bibfnamefont {D.}~\bibnamefont {Charalambidis}},\ }\href@noop {} {\bibfield
   {journal} {\bibinfo  {journal} {Nat. Phys.}\ }\textbf {\bibinfo {volume}
  {3}},\ \bibinfo {pages} {846} (\bibinfo {year} {2007})}\BibitemShut {NoStop}%
\bibitem [{\citenamefont {Feng}\ \emph {et~al.}(2009)\citenamefont {Feng},
  \citenamefont {Gilbertson}, \citenamefont {Mashiko}, \citenamefont {Wang},
  \citenamefont {Khan}, \citenamefont {Chini}, \citenamefont {Wu},
  \citenamefont {Zhao},\ and\ \citenamefont {Chang}}]{Ximao09}%
  \BibitemOpen
  \bibfield  {author} {\bibinfo {author} {\bibfnamefont {X.}~\bibnamefont
  {Feng}}, \bibinfo {author} {\bibfnamefont {S.}~\bibnamefont {Gilbertson}},
  \bibinfo {author} {\bibfnamefont {H.}~\bibnamefont {Mashiko}}, \bibinfo
  {author} {\bibfnamefont {H.}~\bibnamefont {Wang}}, \bibinfo {author}
  {\bibfnamefont {S.~D.}\ \bibnamefont {Khan}}, \bibinfo {author}
  {\bibfnamefont {M.}~\bibnamefont {Chini}}, \bibinfo {author} {\bibfnamefont
  {Y.}~\bibnamefont {Wu}}, \bibinfo {author} {\bibfnamefont {K.}~\bibnamefont
  {Zhao}}, \ and\ \bibinfo {author} {\bibfnamefont {Z.}~\bibnamefont {Chang}},\
  }\href@noop {} {\bibfield  {journal} {\bibinfo  {journal} {Phys. Rev. Lett.}\
  }\textbf {\bibinfo {volume} {103}},\ \bibinfo {pages} {183901} (\bibinfo
  {year} {2009})}\BibitemShut {NoStop}%
\bibitem [{\citenamefont {Skantzakis}\ \emph {et~al.}(2009)\citenamefont
  {Skantzakis}, \citenamefont {Tzallas}, \citenamefont {Kruse}, \citenamefont
  {Kalpouzos},\ and\ \citenamefont {Charalambidis}}]{Skantzakis09}%
  \BibitemOpen
  \bibfield  {author} {\bibinfo {author} {\bibfnamefont {E.}~\bibnamefont
  {Skantzakis}}, \bibinfo {author} {\bibfnamefont {P.}~\bibnamefont {Tzallas}},
  \bibinfo {author} {\bibfnamefont {J.}~\bibnamefont {Kruse}}, \bibinfo
  {author} {\bibfnamefont {C.}~\bibnamefont {Kalpouzos}}, \ and\ \bibinfo
  {author} {\bibfnamefont {D.}~\bibnamefont {Charalambidis}},\ }\href@noop {}
  {\bibfield  {journal} {\bibinfo  {journal} {Opt. Lett.}\ }\textbf {\bibinfo
  {volume} {34}},\ \bibinfo {pages} {1732} (\bibinfo {year}
  {2009})}\BibitemShut {NoStop}%
\bibitem [{\citenamefont {H}, \citenamefont {K},\ and\ \citenamefont
  {T}(2013)}]{Mashiko13}%
  \BibitemOpen
  \bibfield  {author} {\bibinfo {author} {\bibfnamefont {M.}~\bibnamefont {H}},
  \bibinfo {author} {\bibfnamefont {O.}~\bibnamefont {K}}, \ and\ \bibinfo
  {author} {\bibfnamefont {S.}~\bibnamefont {T}},\ }\href@noop {} {\bibfield
  {journal} {\bibinfo  {journal} {Appl Phys Lett,}\ }\textbf {\bibinfo {volume}
  {102}},\ \bibinfo {pages} {171111} (\bibinfo {year} {2013})}\BibitemShut
  {NoStop}%
\bibitem [{\citenamefont {Huang}\ \emph {et~al.}(2011)\citenamefont {Huang},
  \citenamefont {Cirmi}, \citenamefont {Moses}, \citenamefont {Hong},
  \citenamefont {Bhardwaj}, \citenamefont {Birge}, \citenamefont {Chen},
  \citenamefont {Li}, \citenamefont {Eggleton}, \citenamefont {Cerullo},\ and\
  \citenamefont {K\"{o}rtner}}]{Huang11}%
  \BibitemOpen
  \bibfield  {author} {\bibinfo {author} {\bibfnamefont {S.-W.}\ \bibnamefont
  {Huang}}, \bibinfo {author} {\bibfnamefont {G.}~\bibnamefont {Cirmi}},
  \bibinfo {author} {\bibfnamefont {J.}~\bibnamefont {Moses}}, \bibinfo
  {author} {\bibfnamefont {K.-H.}\ \bibnamefont {Hong}}, \bibinfo {author}
  {\bibfnamefont {S.}~\bibnamefont {Bhardwaj}}, \bibinfo {author}
  {\bibfnamefont {J.~R.}\ \bibnamefont {Birge}}, \bibinfo {author}
  {\bibfnamefont {L.-J.}\ \bibnamefont {Chen}}, \bibinfo {author}
  {\bibfnamefont {E.}~\bibnamefont {Li}}, \bibinfo {author} {\bibfnamefont
  {B.~J.}\ \bibnamefont {Eggleton}}, \bibinfo {author} {\bibfnamefont
  {G.}~\bibnamefont {Cerullo}}, \ and\ \bibinfo {author} {\bibfnamefont
  {F.~X.}\ \bibnamefont {K\"{o}rtner}},\ }\href@noop {} {\bibfield  {journal}
  {\bibinfo  {journal} {Nat. Photonics}\ }\textbf {\bibinfo {volume} {5}},\
  \bibinfo {pages} {475} (\bibinfo {year} {2011})}\BibitemShut {NoStop}%
\bibitem [{\citenamefont {Hassan}\ \emph {et~al.}(2016)\citenamefont {Hassan},
  \citenamefont {Luu1}, \citenamefont {Moulet}, \citenamefont {Raskazovskaya},
  \citenamefont {Zhokhov}, \citenamefont {Garg}, \citenamefont {Karpowicz},
  \citenamefont {A.~M.~Zheltikov3}, \citenamefont {Krausz},\ and\ \citenamefont
  {Goulielmakis}}]{Hassan16}%
  \BibitemOpen
  \bibfield  {author} {\bibinfo {author} {\bibfnamefont {M.~T.}\ \bibnamefont
  {Hassan}}, \bibinfo {author} {\bibfnamefont {T.~T.}\ \bibnamefont {Luu1}},
  \bibinfo {author} {\bibfnamefont {A.}~\bibnamefont {Moulet}}, \bibinfo
  {author} {\bibfnamefont {O.}~\bibnamefont {Raskazovskaya}}, \bibinfo {author}
  {\bibfnamefont {P.}~\bibnamefont {Zhokhov}}, \bibinfo {author} {\bibfnamefont
  {M.}~\bibnamefont {Garg}}, \bibinfo {author} {\bibfnamefont {N.}~\bibnamefont
  {Karpowicz}}, \bibinfo {author} {\bibfnamefont {V.~P.}\ \bibnamefont
  {A.~M.~Zheltikov3}, \bibfnamefont {4}}, \bibinfo {author} {\bibfnamefont
  {F.}~\bibnamefont {Krausz}}, \ and\ \bibinfo {author} {\bibfnamefont {.~E.}\
  \bibnamefont {Goulielmakis}},\ }\href@noop {} {\bibfield  {journal} {\bibinfo
   {journal} {Nature}\ }\textbf {\bibinfo {volume} {530}},\ \bibinfo {pages}
  {66} (\bibinfo {year} {2016})}\BibitemShut {NoStop}%
\bibitem [{\citenamefont {Rossi}\ \emph {et~al.}(2020)\citenamefont {Rossi},
  \citenamefont {Mainz}, \citenamefont {Yang}, \citenamefont {Scheiba},
  \citenamefont {Silva-Toledo}, \citenamefont {Chia}, \citenamefont {Keathley},
  \citenamefont {Fang}, \citenamefont {M\"{u}cke}, \citenamefont {Manzoni},
  \citenamefont {Cerullo}, \citenamefont {Cirmi},\ and\ \citenamefont
  {K\"{o}rtner}}]{Giulio20}%
  \BibitemOpen
  \bibfield  {author} {\bibinfo {author} {\bibfnamefont {G.~M.}\ \bibnamefont
  {Rossi}}, \bibinfo {author} {\bibfnamefont {R.~E.}\ \bibnamefont {Mainz}},
  \bibinfo {author} {\bibfnamefont {Y.}~\bibnamefont {Yang}}, \bibinfo {author}
  {\bibfnamefont {F.}~\bibnamefont {Scheiba}}, \bibinfo {author} {\bibfnamefont
  {M.~A.}\ \bibnamefont {Silva-Toledo}}, \bibinfo {author} {\bibfnamefont
  {S.-H.}\ \bibnamefont {Chia}}, \bibinfo {author} {\bibfnamefont {P.~D.}\
  \bibnamefont {Keathley}}, \bibinfo {author} {\bibfnamefont {S.}~\bibnamefont
  {Fang}}, \bibinfo {author} {\bibfnamefont {O.~D.}\ \bibnamefont {M\"{u}cke}},
  \bibinfo {author} {\bibfnamefont {C.}~\bibnamefont {Manzoni}}, \bibinfo
  {author} {\bibfnamefont {G.}~\bibnamefont {Cerullo}}, \bibinfo {author}
  {\bibfnamefont {G.}~\bibnamefont {Cirmi}}, \ and\ \bibinfo {author}
  {\bibfnamefont {F.~X.}\ \bibnamefont {K\"{o}rtner}},\ }\href@noop {}
  {\bibfield  {journal} {\bibinfo  {journal} {Nat. Photonics}\ }\textbf
  {\bibinfo {volume} {14}},\ \bibinfo {pages} {629} (\bibinfo {year}
  {2020})}\BibitemShut {NoStop}%
\bibitem [{\citenamefont {Malkin}, \citenamefont {Shvets},\ and\ \citenamefont
  {Fisch}(1999)}]{Malkin991}%
  \BibitemOpen
  \bibfield  {author} {\bibinfo {author} {\bibfnamefont {V.~M.}\ \bibnamefont
  {Malkin}}, \bibinfo {author} {\bibfnamefont {G.}~\bibnamefont {Shvets}}, \
  and\ \bibinfo {author} {\bibfnamefont {N.~J.}\ \bibnamefont {Fisch}},\
  }\href@noop {} {\bibfield  {journal} {\bibinfo  {journal} {Phys. Rev. Lett}\
  }\textbf {\bibinfo {volume} {82}},\ \bibinfo {pages} {4448} (\bibinfo {year}
  {1999})}\BibitemShut {NoStop}%
\bibitem [{\citenamefont {Malkin}, \citenamefont {Shvets},\ and\ \citenamefont
  {Fisch}(2000)}]{Malkin00}%
  \BibitemOpen
  \bibfield  {author} {\bibinfo {author} {\bibfnamefont {V.~M.}\ \bibnamefont
  {Malkin}}, \bibinfo {author} {\bibfnamefont {G.}~\bibnamefont {Shvets}}, \
  and\ \bibinfo {author} {\bibfnamefont {N.~J.}\ \bibnamefont {Fisch}},\
  }\href@noop {} {\bibfield  {journal} {\bibinfo  {journal} {Phys. Rev. Lett}\
  }\textbf {\bibinfo {volume} {84}},\ \bibinfo {pages} {1208} (\bibinfo {year}
  {2000})}\BibitemShut {NoStop}%
\bibitem [{\citenamefont {Trines}\ \emph {et~al.}(2011)\citenamefont {Trines},
  \citenamefont {Fiuza}, \citenamefont {Bingham}, \citenamefont {Fonseca},
  \citenamefont {Silva}, \citenamefont {Cairns},\ and\ \citenamefont
  {Norreys1}}]{Trines111}%
  \BibitemOpen
  \bibfield  {author} {\bibinfo {author} {\bibfnamefont {R.~M. G.~M.}\
  \bibnamefont {Trines}}, \bibinfo {author} {\bibfnamefont {F.}~\bibnamefont
  {Fiuza}}, \bibinfo {author} {\bibfnamefont {R.}~\bibnamefont {Bingham}},
  \bibinfo {author} {\bibfnamefont {R.~A.}\ \bibnamefont {Fonseca}}, \bibinfo
  {author} {\bibfnamefont {L.~O.}\ \bibnamefont {Silva}}, \bibinfo {author}
  {\bibfnamefont {R.~A.}\ \bibnamefont {Cairns}}, \ and\ \bibinfo {author}
  {\bibfnamefont {P.~A.}\ \bibnamefont {Norreys1}},\ }\href@noop {} {\bibfield
  {journal} {\bibinfo  {journal} {Nat. Phys.}\ }\textbf {\bibinfo {volume}
  {7}},\ \bibinfo {pages} {87} (\bibinfo {year} {2011})}\BibitemShut {NoStop}%
\bibitem [{\citenamefont {Ping}\ \emph {et~al.}(2004)\citenamefont {Ping},
  \citenamefont {Cheng}, \citenamefont {Suckewer}, \citenamefont {Clark},\ and\
  \citenamefont {Fisch}}]{Ping04}%
  \BibitemOpen
  \bibfield  {author} {\bibinfo {author} {\bibfnamefont {Y.}~\bibnamefont
  {Ping}}, \bibinfo {author} {\bibfnamefont {W.}~\bibnamefont {Cheng}},
  \bibinfo {author} {\bibfnamefont {S.}~\bibnamefont {Suckewer}}, \bibinfo
  {author} {\bibfnamefont {D.~S.}\ \bibnamefont {Clark}}, \ and\ \bibinfo
  {author} {\bibfnamefont {N.~J.}\ \bibnamefont {Fisch}},\ }\href@noop {}
  {\bibfield  {journal} {\bibinfo  {journal} {Phys. Rev. Lett}\ }\textbf
  {\bibinfo {volume} {92}},\ \bibinfo {pages} {175007} (\bibinfo {year}
  {2004})}\BibitemShut {NoStop}%
\bibitem [{\citenamefont {Cheng}\ \emph {et~al.}(2005)\citenamefont {Cheng},
  \citenamefont {Avitzour}, \citenamefont {Ping}, \citenamefont {Suckewer},
  \citenamefont {Fisch}, \citenamefont {Hur},\ and\ \citenamefont
  {Wurtele}}]{Cheng05}%
  \BibitemOpen
  \bibfield  {author} {\bibinfo {author} {\bibfnamefont {W.}~\bibnamefont
  {Cheng}}, \bibinfo {author} {\bibfnamefont {Y.}~\bibnamefont {Avitzour}},
  \bibinfo {author} {\bibfnamefont {Y.}~\bibnamefont {Ping}}, \bibinfo {author}
  {\bibfnamefont {S.}~\bibnamefont {Suckewer}}, \bibinfo {author}
  {\bibfnamefont {N.~J.}\ \bibnamefont {Fisch}}, \bibinfo {author}
  {\bibfnamefont {M.~S.}\ \bibnamefont {Hur}}, \ and\ \bibinfo {author}
  {\bibfnamefont {J.~S.}\ \bibnamefont {Wurtele}},\ }\href@noop {} {\bibfield
  {journal} {\bibinfo  {journal} {Phys. Rev. Lett}\ }\textbf {\bibinfo {volume}
  {94}},\ \bibinfo {pages} {045003} (\bibinfo {year} {2005})}\BibitemShut
  {NoStop}%
\bibitem [{\citenamefont {Ren}\ \emph {et~al.}(2007)\citenamefont {Ren},
  \citenamefont {Cheng}, \citenamefont {Li},\ and\ \citenamefont
  {Suckewer}}]{JUN07}%
  \BibitemOpen
  \bibfield  {author} {\bibinfo {author} {\bibfnamefont {J.}~\bibnamefont
  {Ren}}, \bibinfo {author} {\bibfnamefont {W.}~\bibnamefont {Cheng}}, \bibinfo
  {author} {\bibfnamefont {S.}~\bibnamefont {Li}}, \ and\ \bibinfo {author}
  {\bibfnamefont {S.}~\bibnamefont {Suckewer}},\ }\href@noop {} {\bibfield
  {journal} {\bibinfo  {journal} {Nat. Phys.}\ }\textbf {\bibinfo {volume}
  {3}},\ \bibinfo {pages} {732} (\bibinfo {year} {2007})}\BibitemShut {NoStop}%
\bibitem [{\citenamefont {Turnbull}\ \emph {et~al.}(2018)\citenamefont
  {Turnbull}, \citenamefont {Franke}, \citenamefont {Katz}, \citenamefont
  {Palastro}, \citenamefont {Begishev}, \citenamefont {Boni}, \citenamefont
  {Bromage}, \citenamefont {Milder}, \citenamefont {Shaw},\ and\ \citenamefont
  {Froula}}]{Turbunll18s}%
  \BibitemOpen
  \bibfield  {author} {\bibinfo {author} {\bibfnamefont {D.}~\bibnamefont
  {Turnbull}}, \bibinfo {author} {\bibfnamefont {P.}~\bibnamefont {Franke}},
  \bibinfo {author} {\bibfnamefont {J.}~\bibnamefont {Katz}}, \bibinfo {author}
  {\bibfnamefont {J.~P.}\ \bibnamefont {Palastro}}, \bibinfo {author}
  {\bibfnamefont {I.~A.}\ \bibnamefont {Begishev}}, \bibinfo {author}
  {\bibfnamefont {R.}~\bibnamefont {Boni}}, \bibinfo {author} {\bibfnamefont
  {J.}~\bibnamefont {Bromage}}, \bibinfo {author} {\bibfnamefont {A.~L.}\
  \bibnamefont {Milder}}, \bibinfo {author} {\bibfnamefont {J.~L.}\
  \bibnamefont {Shaw}}, \ and\ \bibinfo {author} {\bibfnamefont {D.~H.}\
  \bibnamefont {Froula}},\ }\href@noop {} {\bibfield  {journal} {\bibinfo
  {journal} {Phys. Rev. Lett.}\ }\textbf {\bibinfo {volume} {120}},\ \bibinfo
  {pages} {225001} (\bibinfo {year} {2018})}\BibitemShut {NoStop}%
\bibitem [{\citenamefont {Wu}\ \emph {et~al.}(2019)\citenamefont {Wu},
  \citenamefont {Chen}, \citenamefont {Morozov},\ and\ \citenamefont
  {Suckewer}}]{Wu20}%
  \BibitemOpen
  \bibfield  {author} {\bibinfo {author} {\bibfnamefont {Z.}~\bibnamefont
  {Wu}}, \bibinfo {author} {\bibfnamefont {Q.}~\bibnamefont {Chen}}, \bibinfo
  {author} {\bibfnamefont {A.}~\bibnamefont {Morozov}}, \ and\ \bibinfo
  {author} {\bibfnamefont {S.}~\bibnamefont {Suckewer}},\ }\href@noop {}
  {\bibfield  {journal} {\bibinfo  {journal} {Phys. Plasmas}\ }\textbf
  {\bibinfo {volume} {26}},\ \bibinfo {pages} {103111} (\bibinfo {year}
  {2019})}\BibitemShut {NoStop}%
\bibitem [{\citenamefont {Andreev}\ \emph {et~al.}(2006)\citenamefont
  {Andreev}, \citenamefont {Riconda}, \citenamefont {Tikhonchuk},\ and\
  \citenamefont {Weber}}]{Andreev06}%
  \BibitemOpen
  \bibfield  {author} {\bibinfo {author} {\bibfnamefont {A.~A.}\ \bibnamefont
  {Andreev}}, \bibinfo {author} {\bibfnamefont {C.}~\bibnamefont {Riconda}},
  \bibinfo {author} {\bibfnamefont {V.~T.}\ \bibnamefont {Tikhonchuk}}, \ and\
  \bibinfo {author} {\bibfnamefont {S.}~\bibnamefont {Weber}},\ }\href@noop {}
  {\bibfield  {journal} {\bibinfo  {journal} {Phys. Plasma}\ }\textbf {\bibinfo
  {volume} {13}},\ \bibinfo {pages} {053110} (\bibinfo {year}
  {2006})}\BibitemShut {NoStop}%
\bibitem [{\citenamefont {Weber}\ \emph {et~al.}(2013)\citenamefont {Weber},
  \citenamefont {Riconda}, \citenamefont {Lancia}, \citenamefont {Marque\`{s}},
  \citenamefont {Mourou},\ and\ \citenamefont {Fuchs}}]{Weber13}%
  \BibitemOpen
  \bibfield  {author} {\bibinfo {author} {\bibfnamefont {S.}~\bibnamefont
  {Weber}}, \bibinfo {author} {\bibfnamefont {C.}~\bibnamefont {Riconda}},
  \bibinfo {author} {\bibfnamefont {L.}~\bibnamefont {Lancia}}, \bibinfo
  {author} {\bibfnamefont {J.-R.}\ \bibnamefont {Marque\`{s}}}, \bibinfo
  {author} {\bibfnamefont {G.~A.}\ \bibnamefont {Mourou}}, \ and\ \bibinfo
  {author} {\bibfnamefont {J.}~\bibnamefont {Fuchs}},\ }\href@noop {}
  {\bibfield  {journal} {\bibinfo  {journal} {Phys. Rev. Lett.}\ }\textbf
  {\bibinfo {volume} {111}},\ \bibinfo {pages} {055004} (\bibinfo {year}
  {2013})}\BibitemShut {NoStop}%
\bibitem [{\citenamefont {Lancia}\ \emph {et~al.}(2010)\citenamefont {Lancia},
  \citenamefont {Marqu\`{e}s}, \citenamefont {Nakatsutsumi}, \citenamefont
  {Riconda}, \citenamefont {Weber}, \citenamefont {H\"{u}ller}, \citenamefont
  {A.Man\v{c}i\'{c}}, \citenamefont {Antici}, \citenamefont {Tikhonchuk},
  \citenamefont {H\'{e}ron}, \citenamefont {Audebert},\ and\ \citenamefont
  {Fuchs}}]{Lancia13}%
  \BibitemOpen
  \bibfield  {author} {\bibinfo {author} {\bibfnamefont {L.}~\bibnamefont
  {Lancia}}, \bibinfo {author} {\bibfnamefont {J.~R.}\ \bibnamefont
  {Marqu\`{e}s}}, \bibinfo {author} {\bibfnamefont {M.}~\bibnamefont
  {Nakatsutsumi}}, \bibinfo {author} {\bibfnamefont {C.}~\bibnamefont
  {Riconda}}, \bibinfo {author} {\bibfnamefont {S.}~\bibnamefont {Weber}},
  \bibinfo {author} {\bibfnamefont {S.}~\bibnamefont {H\"{u}ller}}, \bibinfo
  {author} {\bibnamefont {A.Man\v{c}i\'{c}}}, \bibinfo {author} {\bibfnamefont
  {P.}~\bibnamefont {Antici}}, \bibinfo {author} {\bibfnamefont {V.~T.}\
  \bibnamefont {Tikhonchuk}}, \bibinfo {author} {\bibfnamefont
  {A.}~\bibnamefont {H\'{e}ron}}, \bibinfo {author} {\bibfnamefont
  {P.}~\bibnamefont {Audebert}}, \ and\ \bibinfo {author} {\bibfnamefont
  {J.}~\bibnamefont {Fuchs}},\ }\href@noop {} {\bibfield  {journal} {\bibinfo
  {journal} {Phys. Rev. Lett.}\ }\textbf {\bibinfo {volume} {104}},\ \bibinfo
  {pages} {025001} (\bibinfo {year} {2010})}\BibitemShut {NoStop}%
\bibitem [{\citenamefont {Lancia}\ \emph {et~al.}(2016)\citenamefont {Lancia},
  \citenamefont {Giribono}, \citenamefont {Vassura}, \citenamefont
  {Chiaramello}, \citenamefont {Riconda}, \citenamefont {Weber}, \citenamefont
  {Castan}, \citenamefont {Chatelain}, \citenamefont {Frank}, \citenamefont
  {Gangolf}, \citenamefont {Quinn}, \citenamefont {Fuchs},\ and\ \citenamefont
  {Marqu\`{e}s}}]{Lancia16}%
  \BibitemOpen
  \bibfield  {author} {\bibinfo {author} {\bibfnamefont {L.}~\bibnamefont
  {Lancia}}, \bibinfo {author} {\bibfnamefont {A.}~\bibnamefont {Giribono}},
  \bibinfo {author} {\bibfnamefont {L.}~\bibnamefont {Vassura}}, \bibinfo
  {author} {\bibfnamefont {M.}~\bibnamefont {Chiaramello}}, \bibinfo {author}
  {\bibfnamefont {C.}~\bibnamefont {Riconda}}, \bibinfo {author} {\bibfnamefont
  {S.}~\bibnamefont {Weber}}, \bibinfo {author} {\bibfnamefont
  {A.}~\bibnamefont {Castan}}, \bibinfo {author} {\bibfnamefont
  {A.}~\bibnamefont {Chatelain}}, \bibinfo {author} {\bibfnamefont
  {A.}~\bibnamefont {Frank}}, \bibinfo {author} {\bibfnamefont
  {T.}~\bibnamefont {Gangolf}}, \bibinfo {author} {\bibfnamefont {M.~N.}\
  \bibnamefont {Quinn}}, \bibinfo {author} {\bibfnamefont {J.}~\bibnamefont
  {Fuchs}}, \ and\ \bibinfo {author} {\bibfnamefont {J.-R.}\ \bibnamefont
  {Marqu\`{e}s}},\ }\href@noop {} {\bibfield  {journal} {\bibinfo  {journal}
  {Phys. Rev. Lett.}\ }\textbf {\bibinfo {volume} {116}},\ \bibinfo {pages}
  {075001} (\bibinfo {year} {2016})}\BibitemShut {NoStop}%
\bibitem [{\citenamefont {Marqu\'{e}s}\ \emph {et~al.}(2019)\citenamefont
  {Marqu\'{e}s}, \citenamefont {Lancia}, \citenamefont {Gangolf}, \citenamefont
  {Blecher}, \citenamefont {Bolanos}, \citenamefont {J.Fuchs}, \citenamefont
  {Willi}, \citenamefont {Amiranoff}, \citenamefont {Berger}, \citenamefont
  {Chiaramello}, \citenamefont {Weber},\ and\ \citenamefont
  {Riconda}}]{Marques19}%
  \BibitemOpen
  \bibfield  {author} {\bibinfo {author} {\bibfnamefont {J.-R.}\ \bibnamefont
  {Marqu\'{e}s}}, \bibinfo {author} {\bibfnamefont {L.}~\bibnamefont {Lancia}},
  \bibinfo {author} {\bibfnamefont {T.}~\bibnamefont {Gangolf}}, \bibinfo
  {author} {\bibfnamefont {M.}~\bibnamefont {Blecher}}, \bibinfo {author}
  {\bibfnamefont {S.}~\bibnamefont {Bolanos}}, \bibinfo {author} {\bibnamefont
  {J.Fuchs}}, \bibinfo {author} {\bibfnamefont {O.}~\bibnamefont {Willi}},
  \bibinfo {author} {\bibfnamefont {F.}~\bibnamefont {Amiranoff}}, \bibinfo
  {author} {\bibfnamefont {R.~L.}\ \bibnamefont {Berger}}, \bibinfo {author}
  {\bibfnamefont {M.}~\bibnamefont {Chiaramello}}, \bibinfo {author}
  {\bibfnamefont {S.}~\bibnamefont {Weber}}, \ and\ \bibinfo {author}
  {\bibfnamefont {C.}~\bibnamefont {Riconda}},\ }\href@noop {} {\bibfield
  {journal} {\bibinfo  {journal} {Phys. Rev. X}\ }\textbf {\bibinfo {volume}
  {9}},\ \bibinfo {pages} {021008} (\bibinfo {year} {2019})}\BibitemShut
  {NoStop}%
\bibitem [{\citenamefont {Wu}\ \emph {et~al.}(2022)\citenamefont {Wu},
  \citenamefont {Zeng}, \citenamefont {Li}, \citenamefont {Zhang},
  \citenamefont {Wang}, \citenamefont {Hu}, \citenamefont {Wang}, \citenamefont
  {Mu}, \citenamefont {Su}, \citenamefont {Zhu}, \citenamefont {Wei}, ,\ and\
  \citenamefont {Zuo}}]{Zhaohui22}%
  \BibitemOpen
  \bibfield  {author} {\bibinfo {author} {\bibfnamefont {Z.}~\bibnamefont
  {Wu}}, \bibinfo {author} {\bibfnamefont {X.}~\bibnamefont {Zeng}}, \bibinfo
  {author} {\bibfnamefont {Z.}~\bibnamefont {Li}}, \bibinfo {author}
  {\bibfnamefont {Z.}~\bibnamefont {Zhang}}, \bibinfo {author} {\bibfnamefont
  {X.}~\bibnamefont {Wang}}, \bibinfo {author} {\bibfnamefont {B.}~\bibnamefont
  {Hu}}, \bibinfo {author} {\bibfnamefont {X.}~\bibnamefont {Wang}}, \bibinfo
  {author} {\bibfnamefont {J.}~\bibnamefont {Mu}}, \bibinfo {author}
  {\bibfnamefont {J.}~\bibnamefont {Su}}, \bibinfo {author} {\bibfnamefont
  {Q.}~\bibnamefont {Zhu}}, \bibinfo {author} {\bibfnamefont {X.}~\bibnamefont
  {Wei}}, , \ and\ \bibinfo {author} {\bibfnamefont {Y.}~\bibnamefont {Zuo}},\
  }\href@noop {} {\bibfield  {journal} {\bibinfo  {journal} {Matter Radiate
  Extreme}\ }\textbf {\bibinfo {volume} {7}},\ \bibinfo {pages} {064402}
  (\bibinfo {year} {2022})}\BibitemShut {NoStop}%
\bibitem [{\citenamefont {Li}\ \emph {et~al.}(2022)\citenamefont {Li},
  \citenamefont {Zuo}, \citenamefont {Zeng}, \citenamefont {Wu}, \citenamefont
  {Wang}, \citenamefont {Wang}, \citenamefont {Mu},\ and\ \citenamefont
  {Hu}}]{Zhaoli22}%
  \BibitemOpen
  \bibfield  {author} {\bibinfo {author} {\bibfnamefont {Z.}~\bibnamefont
  {Li}}, \bibinfo {author} {\bibfnamefont {Y.}~\bibnamefont {Zuo}}, \bibinfo
  {author} {\bibfnamefont {X.}~\bibnamefont {Zeng}}, \bibinfo {author}
  {\bibfnamefont {Z.}~\bibnamefont {Wu}}, \bibinfo {author} {\bibfnamefont
  {X.}~\bibnamefont {Wang}}, \bibinfo {author} {\bibfnamefont {X.}~\bibnamefont
  {Wang}}, \bibinfo {author} {\bibfnamefont {J.}~\bibnamefont {Mu}}, \ and\
  \bibinfo {author} {\bibfnamefont {B.}~\bibnamefont {Hu}},\ }\href@noop {}
  {\bibfield  {journal} {\bibinfo  {journal} {Matter Radiate Extreme}\ }\textbf
  {\bibinfo {volume} {8}},\ \bibinfo {pages} {014401} (\bibinfo {year}
  {2022})}\BibitemShut {NoStop}%
\bibitem [{\citenamefont {Damzen}\ \emph {et~al.}(2003)\citenamefont {Damzen},
  \citenamefont {Vlad}, \citenamefont {Babin},\ and\ \citenamefont
  {Mocofanescu}}]{SBS}%
  \BibitemOpen
  \bibfield  {author} {\bibinfo {author} {\bibfnamefont {M.}~\bibnamefont
  {Damzen}}, \bibinfo {author} {\bibfnamefont {V.}~\bibnamefont {Vlad}},
  \bibinfo {author} {\bibfnamefont {V.}~\bibnamefont {Babin}}, \ and\ \bibinfo
  {author} {\bibfnamefont {A.}~\bibnamefont {Mocofanescu}},\ }\href@noop {}
  {\emph {\bibinfo {title} {Stimulated Brillouin Scattering Fundamentals and
  Applications}}}\ (\bibinfo  {publisher} {Institute of Physics Publishing},\
  \bibinfo {address} {London},\ \bibinfo {year} {2003})\BibitemShut {NoStop}%
\bibitem [{\citenamefont {Gorbunov}\ \emph {et~al.}(1983)\citenamefont
  {Gorbunov}, \citenamefont {Papernyi}, \citenamefont {Petrov},\ and\
  \citenamefont {Startsev}}]{Gorbunov83}%
  \BibitemOpen
  \bibfield  {author} {\bibinfo {author} {\bibfnamefont {V.~A.}\ \bibnamefont
  {Gorbunov}}, \bibinfo {author} {\bibfnamefont {S.~B.}\ \bibnamefont
  {Papernyi}}, \bibinfo {author} {\bibfnamefont {V.~F.}\ \bibnamefont
  {Petrov}}, \ and\ \bibinfo {author} {\bibfnamefont {V.~R.}\ \bibnamefont
  {Startsev}},\ }\href@noop {} {\bibfield  {journal} {\bibinfo  {journal} {Sov.
  Quantum Electronics}\ }\textbf {\bibinfo {volume} {13}},\ \bibinfo {pages}
  {900} (\bibinfo {year} {1983})}\BibitemShut {NoStop}%
\bibitem [{\citenamefont {Yang}, \citenamefont {Gygerand},\ and\ \citenamefont
  {Th\'{e}venaz}(2020)}]{yang20}%
  \BibitemOpen
  \bibfield  {author} {\bibinfo {author} {\bibfnamefont {F.}~\bibnamefont
  {Yang}}, \bibinfo {author} {\bibfnamefont {F.}~\bibnamefont {Gygerand}}, \
  and\ \bibinfo {author} {\bibfnamefont {L.}~\bibnamefont {Th\'{e}venaz}},\
  }\href@noop {} {\bibfield  {journal} {\bibinfo  {journal} {Nat. Photonics}\
  }\textbf {\bibinfo {volume} {14}},\ \bibinfo {pages} {700} (\bibinfo {year}
  {2020})}\BibitemShut {NoStop}%
\bibitem [{\citenamefont {Carr}\ and\ \citenamefont {Hanna}(1985)}]{Carr85}%
  \BibitemOpen
  \bibfield  {author} {\bibinfo {author} {\bibfnamefont {I.~D.}\ \bibnamefont
  {Carr}}\ and\ \bibinfo {author} {\bibfnamefont {D.~C.}\ \bibnamefont
  {Hanna}},\ }\href@noop {} {\bibfield  {journal} {\bibinfo  {journal} {Applied
  Physics B: Lasers and Optics}\ }\textbf {\bibinfo {volume} {36}},\ \bibinfo
  {pages} {83} (\bibinfo {year} {1985})}\BibitemShut {NoStop}%
\bibitem [{\citenamefont {Meng}\ and\ \citenamefont
  {H.J.Eichler}(1991)}]{Hui91}%
  \BibitemOpen
  \bibfield  {author} {\bibinfo {author} {\bibfnamefont {H.}~\bibnamefont
  {Meng}}\ and\ \bibinfo {author} {\bibnamefont {H.J.Eichler}},\ }\href@noop {}
  {\bibfield  {journal} {\bibinfo  {journal} {Opt. Lett.}\ }\textbf {\bibinfo
  {volume} {16}},\ \bibinfo {pages} {569} (\bibinfo {year} {1991})}\BibitemShut
  {NoStop}%
\bibitem [{\citenamefont {Zhang}\ \emph {et~al.}(2021)\citenamefont {Zhang},
  \citenamefont {Nie}, \citenamefont {Wu}, \citenamefont {Sinclair},
  \citenamefont {Huang}, \citenamefont {Marsh},\ and\ \citenamefont
  {Joshi}}]{chaojie21}%
  \BibitemOpen
  \bibfield  {author} {\bibinfo {author} {\bibfnamefont {C.}~\bibnamefont
  {Zhang}}, \bibinfo {author} {\bibfnamefont {Z.}~\bibnamefont {Nie}}, \bibinfo
  {author} {\bibfnamefont {Y.}~\bibnamefont {Wu}}, \bibinfo {author}
  {\bibfnamefont {M.}~\bibnamefont {Sinclair}}, \bibinfo {author}
  {\bibfnamefont {C.-K.}\ \bibnamefont {Huang}}, \bibinfo {author}
  {\bibfnamefont {K.~A.}\ \bibnamefont {Marsh}}, \ and\ \bibinfo {author}
  {\bibfnamefont {C.}~\bibnamefont {Joshi}},\ }\href@noop {} {\bibfield
  {journal} {\bibinfo  {journal} {Plasma Physics and Controlled Fusion}\
  }\textbf {\bibinfo {volume} {63}},\ \bibinfo {pages} {095011} (\bibinfo
  {year} {2021})}\BibitemShut {NoStop}%
\bibitem [{\citenamefont {Qu}\ \emph {et~al.}(2018)\citenamefont {Qu},
  \citenamefont {Jia}, \citenamefont {Edwards},\ and\ \citenamefont
  {Fisch}}]{Kenan18}%
  \BibitemOpen
  \bibfield  {author} {\bibinfo {author} {\bibfnamefont {K.}~\bibnamefont
  {Qu}}, \bibinfo {author} {\bibfnamefont {Q.}~\bibnamefont {Jia}}, \bibinfo
  {author} {\bibfnamefont {M.~R.}\ \bibnamefont {Edwards}}, \ and\ \bibinfo
  {author} {\bibfnamefont {N.~J.}\ \bibnamefont {Fisch}},\ }\href@noop {}
  {\bibfield  {journal} {\bibinfo  {journal} {Phys. Rev. E}\ }\textbf {\bibinfo
  {volume} {98}},\ \bibinfo {pages} {023202} (\bibinfo {year}
  {2018})}\BibitemShut {NoStop}%
\bibitem [{\citenamefont {K.Qu}\ and\ \citenamefont {Fisch}(2019)}]{Kenan19}%
  \BibitemOpen
  \bibfield  {author} {\bibinfo {author} {\bibnamefont {K.Qu}}\ and\ \bibinfo
  {author} {\bibfnamefont {N.~J.}\ \bibnamefont {Fisch}},\ }\href@noop {}
  {\bibfield  {journal} {\bibinfo  {journal} {Phys. Rev. E}\ }\textbf {\bibinfo
  {volume} {99}},\ \bibinfo {pages} {063201} (\bibinfo {year}
  {2019})}\BibitemShut {NoStop}%
\bibitem [{\citenamefont {Howard}\ \emph {et~al.}(2019)\citenamefont {Howard},
  \citenamefont {Turnbull}, \citenamefont {Davies}, \citenamefont {Franke},
  \citenamefont {Froula},\ and\ \citenamefont {Palastro}}]{Howard19}%
  \BibitemOpen
  \bibfield  {author} {\bibinfo {author} {\bibfnamefont {A.~J.}\ \bibnamefont
  {Howard}}, \bibinfo {author} {\bibfnamefont {D.}~\bibnamefont {Turnbull}},
  \bibinfo {author} {\bibfnamefont {A.~S.}\ \bibnamefont {Davies}}, \bibinfo
  {author} {\bibfnamefont {P.}~\bibnamefont {Franke}}, \bibinfo {author}
  {\bibfnamefont {D.~H.}\ \bibnamefont {Froula}}, \ and\ \bibinfo {author}
  {\bibfnamefont {J.~P.}\ \bibnamefont {Palastro}},\ }\href@noop {} {\bibfield
  {journal} {\bibinfo  {journal} {Phys. Rev. Lett.}\ }\textbf {\bibinfo
  {volume} {123}},\ \bibinfo {pages} {124801} (\bibinfo {year}
  {2019})}\BibitemShut {NoStop}%
\bibitem [{\citenamefont {Esarey}, \citenamefont {G.Joyce},\ and\ \citenamefont
  {Sprangle}(1991)}]{Esarey91}%
  \BibitemOpen
  \bibfield  {author} {\bibinfo {author} {\bibfnamefont {E.}~\bibnamefont
  {Esarey}}, \bibinfo {author} {\bibnamefont {G.Joyce}}, \ and\ \bibinfo
  {author} {\bibfnamefont {P.}~\bibnamefont {Sprangle}},\ }\href@noop {}
  {\bibfield  {journal} {\bibinfo  {journal} {Phys. Rev. A}\ }\textbf {\bibinfo
  {volume} {44}},\ \bibinfo {pages} {3908} (\bibinfo {year}
  {1991})}\BibitemShut {NoStop}%
\bibitem [{\citenamefont {H.Peng}\ \emph {et~al.}(2021)\citenamefont {H.Peng},
  \citenamefont {Riconda}, \citenamefont {S.Weber}, \citenamefont {Zhou},\ and\
  \citenamefont {Ruan}}]{Peng21}%
  \BibitemOpen
  \bibfield  {author} {\bibinfo {author} {\bibnamefont {H.Peng}}, \bibinfo
  {author} {\bibfnamefont {C.}~\bibnamefont {Riconda}}, \bibinfo {author}
  {\bibnamefont {S.Weber}}, \bibinfo {author} {\bibfnamefont {C.~T.}\
  \bibnamefont {Zhou}}, \ and\ \bibinfo {author} {\bibfnamefont {S.~C.}\
  \bibnamefont {Ruan}},\ }\href@noop {} {\bibfield  {journal} {\bibinfo
  {journal} {Phys. Rev. Applied}\ }\textbf {\bibinfo {volume} {15}},\ \bibinfo
  {pages} {054053} (\bibinfo {year} {2021})}\BibitemShut {NoStop}%
\bibitem [{\citenamefont {Arber}\ \emph {et~al.}(2015)\citenamefont {Arber},
  \citenamefont {Bennett}, \citenamefont {Brady}, \citenamefont
  {Lawrence-Douglas}, \citenamefont {Ramsay}, \citenamefont {Sircombe},
  \citenamefont {Gillies}, \citenamefont {Evans}, \citenamefont {Schmitz},
  \citenamefont {Bell},\ and\ \citenamefont {Ridgers}}]{Arber15}%
  \BibitemOpen
  \bibfield  {author} {\bibinfo {author} {\bibfnamefont {T.~D.}\ \bibnamefont
  {Arber}}, \bibinfo {author} {\bibfnamefont {K.}~\bibnamefont {Bennett}},
  \bibinfo {author} {\bibfnamefont {C.~S.}\ \bibnamefont {Brady}}, \bibinfo
  {author} {\bibfnamefont {A.}~\bibnamefont {Lawrence-Douglas}}, \bibinfo
  {author} {\bibfnamefont {M.~G.}\ \bibnamefont {Ramsay}}, \bibinfo {author}
  {\bibfnamefont {N.~J.}\ \bibnamefont {Sircombe}}, \bibinfo {author}
  {\bibfnamefont {P.}~\bibnamefont {Gillies}}, \bibinfo {author} {\bibfnamefont
  {R.~G.}\ \bibnamefont {Evans}}, \bibinfo {author} {\bibfnamefont
  {H.}~\bibnamefont {Schmitz}}, \bibinfo {author} {\bibfnamefont {A.~R.}\
  \bibnamefont {Bell}}, \ and\ \bibinfo {author} {\bibfnamefont {C.~P.}\
  \bibnamefont {Ridgers}},\ }\href@noop {} {\bibfield  {journal} {\bibinfo
  {journal} {Nsma Phys. Controlled Fusion}\ }\textbf {\bibinfo {volume} {57}},\
  \bibinfo {pages} {113001} (\bibinfo {year} {2015})}\BibitemShut {NoStop}%
\bibitem [{\citenamefont {Ammosov}, \citenamefont {Delone},\ and\ \citenamefont
  {Krainov}(1986)}]{Ammosov86}%
  \BibitemOpen
  \bibfield  {author} {\bibinfo {author} {\bibfnamefont {M.}~\bibnamefont
  {Ammosov}}, \bibinfo {author} {\bibfnamefont {N.}~\bibnamefont {Delone}}, \
  and\ \bibinfo {author} {\bibfnamefont {V.}~\bibnamefont {Krainov}},\
  }\href@noop {} {\bibfield  {journal} {\bibinfo  {journal} {Sov. Phys. JETP}\
  }\textbf {\bibinfo {volume} {64}},\ \bibinfo {pages} {1991} (\bibinfo {year}
  {1986})}\BibitemShut {NoStop}%
\bibitem [{\citenamefont {Strickland}\ and\ \citenamefont
  {Mourou}(1985)}]{Mourou85}%
  \BibitemOpen
  \bibfield  {author} {\bibinfo {author} {\bibfnamefont {D.}~\bibnamefont
  {Strickland}}\ and\ \bibinfo {author} {\bibfnamefont {G.}~\bibnamefont
  {Mourou}},\ }\href@noop {} {\bibfield  {journal} {\bibinfo  {journal} {Opt.
  Commun.}\ }\textbf {\bibinfo {volume} {56}},\ \bibinfo {pages} {219}
  (\bibinfo {year} {1985})}\BibitemShut {NoStop}%
\bibitem [{\citenamefont {Mourou}\ \emph {et~al.}(2012)\citenamefont {Mourou},
  \citenamefont {Fisch}, \citenamefont {Malkin}, \citenamefont {Toroker},
  \citenamefont {Khazanov}, \citenamefont {Sergeev}, \citenamefont {Tajima},\
  and\ \citenamefont {Garrec}}]{Mourou12}%
  \BibitemOpen
  \bibfield  {author} {\bibinfo {author} {\bibfnamefont {G.~A.}\ \bibnamefont
  {Mourou}}, \bibinfo {author} {\bibfnamefont {N.~J.}\ \bibnamefont {Fisch}},
  \bibinfo {author} {\bibfnamefont {V.~M.}\ \bibnamefont {Malkin}}, \bibinfo
  {author} {\bibfnamefont {Z.}~\bibnamefont {Toroker}}, \bibinfo {author}
  {\bibfnamefont {E.~A.}\ \bibnamefont {Khazanov}}, \bibinfo {author}
  {\bibfnamefont {A.~M.}\ \bibnamefont {Sergeev}}, \bibinfo {author}
  {\bibfnamefont {T.}~\bibnamefont {Tajima}}, \ and\ \bibinfo {author}
  {\bibfnamefont {B.~L.}\ \bibnamefont {Garrec}},\ }\href@noop {} {\bibfield
  {journal} {\bibinfo  {journal} {Opt. Commun.}\ }\textbf {\bibinfo {volume}
  {285}},\ \bibinfo {pages} {720} (\bibinfo {year} {2012})}\BibitemShut
  {NoStop}%
\end{thebibliography}%

\end{document}